# Surface magnetism of strontium titanate

J. M. D. Coey, M Venkatesan and P Stamenov

School of Physics, Trinity College, Dublin 2, Ireland.

*Abstract.* SrTiO$_3$ plays a central role in oxide electronics. It is the substrate of choice for functional oxide heterostructures based on perovskite-structure thin-film stacks, and its surface or interface with a polar oxide such as LaAlO$_3$ can become a two-dimensional conductor because of electronic reconstruction or the presence of oxygen defects. Inconsistent reports of magnetic order in SrTiO$_3$ abound in the literature. Here we report a systematic experimental study aimed at establishing how and when SrTiO$_3$ can develop a magnetic moment at room temperature. Polished 100, 110 or 111 crystal slices from four different suppliers are characterized before and after vacuum annealing at 750 °C, both in single-crystal and powdered form. Impurity content is analysed at the surface and in the bulk. Besides the underlying intrinsic diamagnetism of SrTiO$_3$, magnetic signals are of three types – a Curie law susceptibility due to dilute magnetic impurities at the ppm level, a hysteretic, temperature-dependent ferromagnetic impurity contribution, and a practically-anhysteretic, defect-related temperature-independent component that saturates in about 200 mT. The latter component is intrinsic. It is often the largest, reaching 10 Bohr magnetons per nm$^2$ of surface area or more and dominating the magnetic response in low fields at room temperature. It is associated with defects near the surface, and can be destroyed by treatment with Tiron (C$_6$H$_4$Na$_2$O$_8$S$_2$), an electron donor molecule that forms a strong complex with titanium at the surface. The origin of this unusual high-temperature ferromagnetic-like response is discussed.

## 1. Introduction.

SrTiO$_3$ (STO) is a transparent insulator with an indirect band gap of 3.2 eV separating the 2$p$(O) valence band and the 3$d$(Ti) $t_{2g}$ conduction band. The oxide has an unusually high dielectric constant of 300 at room temperature, and it becomes conducting in the bulk after annealing in vacuum, when it turns grey. Electron doping with 0.1% Nb is sufficient to make STO a transparent conducting oxide.

The cubic perovskite structure of SrTiO$_3$ shown in Fig 1a) has space group $Pm\underline{3}m$ and lattice parameter $a_0$ = 390.5 pm with Ti octahedrally coordinated by oxygen; the schematic electronic structure is shown in Fig 1b). STO is important for oxide electronics, and it is widely used as a substrate for growing thin films and heterostructures of other functional oxides. 100-cut crystals are nonpolar, and may be TiO$_2$- or SrO-terminated, but 110- or 111-cut crystals are polar with a formal layer charge of ±4e, which leads to surface reconstruction with a layer of tetrahedrally-coordinated titanium [1, 2]. Many other surface reconstructions are possible, depending on surface hydration [3].

The common point defect in STO is the oxygen vacancy [4]. These vacancies can be created by heating the crystals under vacuum, leaving a neutral defect with two associated electrons. An isolated vacancy may be a hydrogenic double donor (The Bohr radius is large on account of the high dielectric constant), but density-functional theory calculations [5-8] show a tendency for one of the electrons to form a σ-bond between a pair of adjacent Ti ions in a hybridized $d_{z2}/4p_z$ orbital state that lies 0.5 – 1.0 eV below the conduction band edge, while the other is delocalized at the bottom of the $t_{2g}$ conduction band. Other configurations with multiple oxygen vacancies and different charges usually lead to localized states with unpaired spin in the gap, and unpolarized or spin polarized electrons at the bottom of the conduction band [6]. These states tend to form small and large polarons, respectively [9]. Oxygen vacancies at the surface may behave similarly [10, 11], or else form pairs normal to the surface, where the upper one contributes two delocalized electrons to the $d_{xy}$ band and the lower one contributes a localized electron to an $e_g$ state of each adjacent Ti ion [12].

A feature that has attracted enormous interest is the interface electronic structure and conductivity that appears when a TiO$_2$-terminated 100 crystal is covered with more than 4 unit cells of a polar perovskite-structure such as LaAlO$_3$ [13-15], which has a formal layer charge of ±e. The effect is attributed either to interface electronic reconstruction to avoid a polar catastrophe, or else to the appearance of oxygen vacancies near the interface [16]. Either way, there is interfacial band bending that leads to the formation of a two-dimensional electron gas (2DEG) at the STO interface; the 2DEG may exhibit high mobility at low temperature. A similar band-bending effect is thought to arise at free $SrTiO_3$ surfaces [17-19] where ARPES measurements show a universal set of broad and narrow bands in the vicinity of the surface, that are independent of the electronic structure of the bulk [18]. Two broad bands with a low effective mass ∼ 0.6 $m_e$ and circular Fermi surfaces correspond to the first two states in the surface quantum well where Ti $d_{xy}$ electrons are delocalized parallel to the interface, but confined in the perpendicular z-direction. Two narrow bands formed from $d_{yz/zx}$ electrons with a high effective mass ∼ 10 - 20 $m_e$ penetrate about two nanometers into the bulk; one is full, and the other partly filled with an ellipsoidal Fermi surface [2, 18-20]. There is a theoretical expectation that electrons in $3d(t_{2g})$ orbitals at the $SrTiO_3$ surface experience a strong Rashba S-O interaction with an enhanced spin splitting of several meV at the points where where the light and heavy bands intersect [10, 21-23].

There are various of reports in the literature of magnetism in STO, and in the interfacial 2DEG [24]. Many of these effects are found only at low temperature. They include Kondo scattering [24-27] and magnetoresistance [24, 28] as well as more direct evidence from magnetometry [29] and stray magnetic field measurements [30]. It is possible to induce small moments that are stable below 10 K and it is thought that oxygen vacancies are needed to induce ferromagnetism in the surface 2DEG [31-33].

There are also reports of room-temperature magnetism. These are based on direct measurements of magnetic hysteresis [34-36], including results following irradiation with an excimer laser [37], and observations by magnetic force microscopy of a gateable stray field [38]. Opposite chiral spin textures are observed with spin winding around the Fermi surface of the first two of the broad $d_{xy}$ sub-bands, which is a consequence of the Rashba spin-orbit coupling [20, 39], and complex, partially-ferromagnetic spin textures have been proposed experimentally [20, 39] and theoretically [10, 11] from the interaction of local moments associated with mid-gap vacancy states and light $d_{xy}$ electrons in the presence of Rashba spin-orbit splitting. However, the claim of a giant Rashba spin splitting of almost 100 meV [20] has neither been confirmed experimentally [39] (possibly due to differences in surface preparation) nor in electronic structure calculations [10]. On the basis of the splitting reported in[20], Gorkov had proposed that STO could be a high-temperature antiferromagnet, on account of a Pomeranchuk instability of the electron gas [40]. There is also a suggestion that ferromagnetism might arise from spontaneous spin splitting of flat $d_{yz/zx}$ bands calculated for (110) STO surfaces [2].

Here our aim was to seek sound evidence of intrinsic high-temperature magnetism in STO. The approach is experimental, taking account of the small signals involved and cognisant of various possible artefacts. Moments reported in SQUID measurements of STO crystal slices with masses of order 40 mg, are small, usually no more than $10^{-8}$ Am$^2$ (10 μemu) and nagging doubts persist that they might be artefacts arising from tiny amounts of ferromagnetic inclusions [41], or other extrinsic sources of contamination [42-47]. We have examined single crystals from four different commercial suppliers, which were subjected to a series of treatments – reduction in vacuum, grinding into powder, physical and chemical surface treatments – together with chemical analysis, to ascertain the origin of the weak magnetic signals observed. The materials are illustrated in Fig 1 c) – e). Preliminary results on crystals from one of these suppliers were reported previously [48]. We are able to separate impurity effects from more interesting physics, specifically an intrinsic high-temperature 'ferromagnetic-like' response at oxygen-deficient STO surfaces.

Our study carries a cautionary message for those working with the 'silicon of oxide electronics' and underlines the variability of STO crystals when considering defects and impurities at the ppm level. It also uncovers an aspect of their magnetism that seems to be quite new.

## 2. Materials.

The study is based on 40 two-side-polished single crystals of STO, of the type often used as substrates for thin film growth, especially by pulsed-laser deposition. The crystal thickness is ≈ 500 μm, and the lateral dimensions are usually either 4 x 4 mm$^2$ or 5 x 5 mm$^2$. The transparent crystals were sourced from four different commercial suppliers, designated as 'A', 'B', 'C', and 'D'. All the samples were handled carefully, to avoid potential sources of contamination. They were never in contact with metal after receipt. The treatments were also carried out in such a way as to minimize the possibility of magnetic contamination. Details are given in the Methods section at the end of the paper.

Magnetization curves $m(H)$ were first measured on all crystals as-received from the suppliers, and then after various treatments, especially heating in vacuum at different temperatures. The vacuum-treated crystals are pale grey. These treatments are relevant because thin films and heterostructures are often grown on $SrTiO_3$ substrates at elevated temperature in vacuum or in a reduced oxygen pressure.

Then all the crystals were studied again after grinding them to powder. The powders were also measured before and after heat treatment in vacuum, mostly at a single temperature in the range 700 – 800 °C. The idea was to increase the surface area by about a factor of 20, in order to see whether the magnetic signals originated from a region close to the surface, or from the bulk. Initially, the powders were white, but they become grey when reduced in vacuum, as shown in Fig. 1d. The vacuum treatments of crystals and powders were intended to introduce oxygen vacancies at the surface.

Most magnetic measurements were made using a 5 T SQUID magnetometer (MPMS 5 XL, Quantum Design) on crystals mounted in clear plastic straws. The lower limit of significance for magnetic moment is chosen, somewhat arbitrarily, as $0.5 \, 10^{-9} \, Am^2$ (0.5 µemu). For comparison, the diamagnetic moment of a 4 x 4 x 0.5 $mm^3$ crystal of $SrTiO_3$ in 5 T is $-210 \, 10^{-9} \, Am^2$. A 4 x 4 $mm^2$ layer of ferromagnetic iron one unit cell thick has a moment of $8 \, 10^{-9} \, Am^2$. The diamagnetic moment in 5 T of the gelcaps used to mount the powders is $-730 \, 10^{-9} \, Am^2$.

## 3. Results.

*Magnetization.* In Fig. 2, we show magnetization curves for one of the $SrTiO_3$ materials, 100 polished crystal slices from supplier 'A', which were measured under the four conditions – A first slice was measured as received, and after annealing in vacuum at 750°C; A second slice was powdered, and then measured before and after annealing in vacuum at 750°C. Raw data are shown in Fig. 2a, and in Fig. 2b they are corrected for a linear diamagnetic background that corresponds to the diamagnetic susceptibility of the crystals, or the sum of the diamagnetic susceptibility and the diamagnetism of the gelcap powder sample holder.

The susceptibility of the 12 crystals at room temperature was -1.36 ± 0.06 $10^{-9}$ m$^3$kg$^{-1}$. There was no appreciable anisotropy, and the diamagnetism was not significantly altered by vacuum annealing or by reducing the crystals to powder.

Differences in the ferromagnetic-like signals show up clearly after subtracting the linear diamagnetic background. For the as-received 4 x 4 mm$^2$ 100 crystal from supplier A, the moment is < 0.5 $10^{-9}$ Am$^2$. This limit corresponds to about 1 $\mu_B$ nm$^{-2}$ of crystal surface (40 mm$^2$). Annealing in vacuum produces no significant change. When another 100 crystal from the same supplier is ground to powder, a significant moment of 2.4 $10^{-9}$ Am$^2$ appears. The largest moment of 39 $10^{-9}$ Am$^2$ is found after the powder is vacuum annealed. It falls to 19 $10^{-9}$ Am$^2$ on re-annealing the powder again at 750 °C in oxygen. No appreciable hysteresis or temperature dependence is associated with the magnetization curve (Fig. 2d). Since the surface area is increased by a factor of roughly 20 when the crystal is reduced to powder, the moment here is approximately 5 $\mu_B$ per nm$^2$.

Crystals from supplier A with the other two orientations behave differently. The 111 crystals show no moment as received, but develop moments of up to 12 $10^{-9}$ Am$^2$ (32 $\mu_B$ nm$^{-2}$) on annealing in vacuum. The as-received 110 crystals already exhibit hysteresis (Fig. 2c), which shows the normal temperature-dependence expected of a ferromagnetic material. This is quite unlike the behavior of the 100 (A) reduced powder in Fig 2b), which shows a saturating magnetization curve that exhibits almost no temperature dependence between 4 K and 380 K, and is practically anhysteretic, $\mu_0 H_c$ < 20 mT (Fig 2d). The main increase of magnetization is again found when the powder is reduced under vacuum. Not surprisingly, the magnetization of the annealed powder is roughly similar, ~ 50 $10^{-9}$ Am$^2$, regardless of the original orientation of the crystal that was ground up to make the powder. This value corresponds to a moment of about 7 $\mu_B$ per nm$^2$, or 1 $\mu_B$ per surface unit cell.

All the specific magnetization σ data on the crystals from supplier 'A' are summarized in Fig. 3a). The corresponding data for SrTiO$_3$ from the other three suppliers are summarized for comparison in Fig. 3b) – d). Note that only two of the 12 crystals are free of ferromagnetic contamination at the 0.5 $10^{-9}$ Am$^2$ level in the as-received state, and that almost all of them develop moments that are two orders of magnitude greater than this value in the reduced powder. Moments of reduced crystals and powders from supplier 'C' are 10 – 12 $\mu_B$ nm$^{-2}$. Crystals from Supplier 'D' show smaller values for the moments of the reduced powder than the others. Here there is less magnetism associated with surface defects, which might be explained by strontium substoichiometry, leading to the presence of some $Ti^{2+}$ on $Sr^{2+}$ sites in the structure.

We illustrate the influence of annealing temperature on the crystals from supplier 'A' in Fig. 4. It was from these data that the annealing temperature of ~ 750 °C was selected. At first we thought that the difference was related to the polarity of the crystal cut [48]; The 100 cut is nonpolar, with no net charge of the alternating SrO and $TiO_2$ planes in the perovskite structure (Fig. 1a). In contrast, the 110 and 111 crystals have charges ±4e. This hypothesis did not stand up, because crystals from the other suppliers do not conform to the pattern. (Fig. 3).

Another aspect is the temperature-dependence of the susceptibility $\chi$, shown for 100 (A) in Fig 2e, where a small Curie-law upturn appears at low temperatures. From the $1/T$ variation it is possible to deduce the product $np_{eff}^2$, where $n$ is the number density of paramagnetic impurities and $p_{eff}$ is the effective Bohr magneton number. For the 100 (A) sample, the upturn would correspond to 0.2 ppm of $Fe^{3+}$, but for 100 (B) it would correspond to 0.9 ppm.

*Chemical analysis.* The magnetic moment data presented in Figs 2 and 3 point to the presence of quite variable quantities of magnetic impurities in the $SrTiO_3$ crystals from different sources, albeit at the ppm level. But they also indicate the creation of some sort of magnetism in $SrTiO_3$ by the heat treatment under vacuum, particularly when fresh surfaces are produced by powdering the crystals. We therefore need to identify the nature and origin of the different signals more precisely. Total impurity content was analysed in the four 100 samples by inductively-coupled plasma mass spectroscopy (ICPMS), and the surfaces were examined by laser ablation mass spectroscopy (LAMS) where rastering a laser spot with diameter 45 μm across the crystal surface allows a spatial resolution of 10 μm.

The main magnetic 3*d* impurities in the $SrTiO_3$ crystals are Fe and Ni. Average concentrations of Fe, Co and Ni in powders made from 100 crystals from the four suppliers are listed in Table 1. For all except 'B', the total content of magnetic 3*d* atoms is less than about I ppm. However, scans like that of Figure 5 show surface concentrations that are much higher than the average, with an erratic spatial distribution of the impurity elements. This suggests that sub-micron inclusions of an iron-nickel phase are present in the surface region, although we were unable to pick them up in high-resolution scanning electron microscopy. The average surface concentrations for the 100 (A) and 110 (A) slices are five times greater than the bulk averages.

Table 1. Chemical Analysis of 3$d$ impurities in 100 crystals (parts per billion)

| Supplier | Fe | Co | Ni |
| --- | --- | --- | --- |
| A | 226 | 10 | 337 |
| B | 5234 | 407 | 32 |
| C | 576 | 21 | 597 |
| D | 98 | 26 | 20 |

*Surface magnetism.* In order to examine further the issue of surface contamination of the polished crystals, a series of experiments was undertaken, aimed at removing 3$d$ impurities in the surface layer. These involved mechanical thinning, molten salt etching and thermal treatment in hot wax. The effect of mechanical thinning by polishing a 110 (A) crystal is illustrated in Figure 6a). Unexpectedly, lapping down one side of the crystal appears to remove almost all of the magnetic signal, not just half of it. The reason was traced to the lapping procedure. The crystal was mounted using hot 'Crystal-Clear' wax (see experimental section). Simply treating one or both sides in this way without polishing was effective at reducing the magnetization of the surface (Fig. 6b). An alternative treatment of the vacuum-annealed 110 (A) substrate in molten NaCl at 800 °C in a Pt crucible was also effective at reducing the magnetic signal of the sample, by 90 % (Fig. 6c). The hot salt is expected to leach 3$d$ impurities from the surfaces, and the crystal turns yellowish after the treatment. Furthermore, we find that it is generally possible to restore some of the magnetic moment that was destroyed by surface treatment with hot wax or molten salt by re-annealing under vacuum.

The large increases of magnetization produced by powdering the crystals and heating the powder under vacuum (Fig 2b) are inexplicable in terms of transition-metal impurities. The 'ferromagnetic' signal from the powder is clearly different from the hysteretic ferromagnetism we have just been considering, insofar as it exhibits little or no hysteresis, and no appreciable temperature dependence, which would suggest a ferromagnetic Curie temperature in excess of 1000 K. The anhysteretic signals can be over ten times greater than could be accounted for by the amounts of impurity that are present. We conclude that this magnetization must somehow be associated with *defects near the surface, or with the surface itself.*

A series of post-grinding treatments of the powder, shown in Fig. 7, confirms that the surface is the origin of the largely-anhysteretic magnetic signal that exhibits no appreciable temperature dependence between 4 K and 380 K (Fig 2d). SrTiO$_3$ is known to darken and become conducting when it is strongly reduced, and our crystals and powders all turn pale grey when they are treated in vacuum at 650 - 750 °C. Optical spectra (not shown) evidence an increased broad-band absorption below the band edge at 3.2 eV. Photoluminescence peaks observed at 2.69, 2.46 and 2.10 eV may be associated with oxygen defect levels in the gap. The enhanced magnetic signal produced by vacuum annealing the powder can be diminished by reannealing in oxygen (Fig 7a). There is also a slight reduction of the magnetic signal after immersion of the powder in a solvent such as ethanol or water and subsequent drying, but we found that the most effective treatment is to disperse the powder in a solution of 'Tiron'

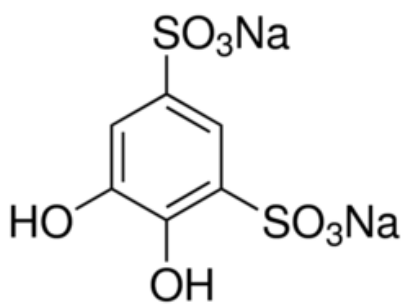

(disodium 4,5-dihydroxy-1,3-benzenedisulfonate) dissolved in ethanol, with a drop of ethylene glycol. Tiron is a superoxide anion scavenger and dispersant that forms strong chemical bonds with titanium, belonging to the catcheol family [49]. It acts as an antioxidant by scavenging free radicals, and transferring electrons to the metal oxide surface. Immersion of a crystal slice in Tiron solution reduces its moment by 20 %, but the moment of a freshly-reduced powder is decreased by 90% [Fig. 7b]. This demonstrates convincingly that the anhysteretic magnetic moment is somehow associated with the electronic structure of surface defects. Treatment with benzoic acid, an electron acceptor, has no such effect.

Finally, we examined a 4 x 4 mm$^2$ 100 (B) crystal of 1.0 wt% Nb-doped SrTiO$_3$, which is conducting because of the addition of Nb *d*-electrons to the Ti 3*d* band. The anhysteretic moment of 10 10$^{-9}$ Am$^2$ is increased by only 10 % when the crystal is reduced to powder. In this case, it appears that the moment does not originate at the SrTiO$_3$ surface, but in the bulk.

The numerical data on crystals and powders are summarized in Table 2.

## 3. Discussion

Since we took precautions in our experiments to avoid contamination of the samples during handling or measurement, the magnetic response can be regarded as characteristic of the SrTiO$_3$ samples themselves. Four components are identified:

* *Intrinsic diamagnetism*. The dimensionless SI susceptibility of all SrTiO$_3$ crystals is -6.5 ± 0.3 10$^{-6}$. The error is one standard deviation, and any variation with crystal orientation is less than 5%. The susceptibility after the crystals are reduced, -6.6 ± 0.4 10$^{-6}$, is not significantly different. There is very little temperature dependence. The diamagnetic susceptibility of the powders is not appreciably different from that of the crystals, although the uncertainly in the measurement is greater due to the dominant contribution of the gelcap sample holders (Fig 2a). The diamagnetic susceptibility of a commercial 5N SrTiO$_3$ powder (Sigma Aldrich) is also similar.

It is expected that the susceptibility of metal oxides containing no cations with unpaired spins should be dominated by the orbital diamagnetism of the oxygen anions (-9.4 10$^{-9}$ m$^3$kg$^{-1}$) plus that of any large cations in the structure [50]. Based on the tabulated ionic contributions, this analysis works quite well for perovskites such as LaAlO$_3$, where the prediction (-21.6 10$^{-6}$) is not far from the experimental value (-18.0 10$^{-6}$). However, the prediction for SrTiO$_3$ (–19.4 10$^{-6}$) disagrees with the measurement by a factor of three. The susceptibility of oxides containing Ti$^{4+}$ is generally greater (less negative) that expected, which is explained by a temperature-independent (Van Vleck) paramagnetic contribution in SrTiO$_3$ of about 12 10$^{-6}$, arising from unoccupied low-lying Ti$^{3+}$ states [51]. The temperature-independent paramagnetism actually outweighs the orbital diamagnetism in rutile TiO$_2$, and the net susceptibility becomes positive [48].

* *Curie-law paramagnetism*. In some samples, this component is comparable in magnitude at 4 K to the intrinsic diamagnetism. Its magnitude varies greatly from sample to sample, and like the intrinsic diamagnetism it does not change appreciably if the crystal is powdered or reduced. The natural explanation is the presence of traces of magnetic transition-metal ions such as Fe$^{3+}$ or Ni$^{2+}$ in the

crystals. The amounts are usually below 1 ppm, and they rarely exceed a few ppm. The Curie-law upturn found in the 5N commercial powder corresponded to 3 ppm of ions with $S = 2$.

* *Extrinsic hysteretic ferromagnetism*. Here the hysteresis increases from 10 mT at RT to 70 mT at 4 K, and there is also an increase in the magnitude of the magnetization (Fig 2c). This is all quite normal for a ferromagnet below its Curie temperature. The magnetization curve measured at room temperature is independent of the direction of the applied field, whether perpendicular or in the plane of the crystal, (Fig.8a). It follows that the demagnetizing factor $\mathcal{N}$ for the ferromagnetic inclusions is the same in all directions, and therefore $\mathcal{N} \approx 1/3$. The magnetization $M_s$ of the ferromagnetic regions can then be estimated as $3H_0$, were $H_0 \approx 80$ kAm$^{-1}$ is deduced by extrapolating the slope of the initial magnetization curve to saturation. Hence $M_s \sim 240$ kAm$^{-1}$, roughly a seventh that of Fe or half the value for Ni, or but close to that for NiFe$_2$O$_4$ (330 kAm$^{-1}$). There is abundant evidence, summarized in Figs 5 and 6, that the ferromagnetic impurity material is present as sparse nuggets in the surface of the crystals. The volume fraction for the unannealed sample of Fig. 2c) can be estimated from the moment of 6 10$^{-9}$ Am$^2$, which gives an average magnetization of 0.75 Am$^{-1}$ over the crystal volume [52]. The hysteretic ferromagnetic volume fraction is therefore 3 ppm. We associate it with the Ni-Fe inclusions that are evident in laser mass spectroscopy scans of Fig. 5. Contamination at this level is likely to be a result of the polishing process, and it reflects the quality of the crystal surface. The signal is largely eliminated by the molten salt treatment, but it is unchanged by powdering.

* *Intrinsic high-temperature surface magnetism*. The fourth component is undoubtedly the most interesting. There is an easily-saturated 'ferromagnetic' signal from SrTiO$_3$, which is somehow associated with defects near the surface. The signal is much increased when new surfaces are created by powdering the crystals (Figs 2b, 3), and it is further enhanced by heating the crystals and especially the powder in vacuum (Figs 2b,3,7a), which often increases the moment by more than an order of magnitude. Conversely, its magnitude is diminished by heating in oxygen (Fig 7a), or by treating the surface with Tiron (Fig 7b). The signal is found in almost all the samples studied, including the 5N powder, and its magnitude can exceed 10 $\mu_B$ nm$^{-2}$, or 1.5 $\mu_B$ per surface unit cell. The field-dependent magnetization curves are independent of temperature in the range 4 – 380 K, and they are practically anhysteretic — coercivity is less than 20 mT at all

temperatures, except for the reoxidized powder in Fig 7a. The extrapolated saturation field $\mu_0 H_0$ is 160 - 220 mT for annealed powders, but measurements on annealed single crystal slices show a higher value, ≈ 400 mT, that is surprisingly independent of the direction of the applied field relative to the crystal surface (Fig 8b,c). This is a truly high-temperature phenomenon, unlike the various reports of Kondo scattering, or low-temperature magnetic ordering in the helium temperature range [24-26, 28-30, 53]. Furthermore, as discussed in the Introduction, the spin moments available in the oxygen-deficient SrTiO$_3$ are only those associated with localized or delocalized electrons with $s = ½$.

Before considering possible explanations of this remarkable surface magnetism — and we stress how unusual it is — we emphasize that the material is not superparamagnetic. Superparamagnets exhibit no hysteresis, but their magnetization curves scale with $H/T$. The initial slope of the magnetization curve in an external field will increase as $1/T$ until it is limited by the demagnetizing field. In the present case, the effect is athermal; there is no temperature dependence at all, and no sign of blocking down to 2 K. (Figs 2d). Magnetically-ordered clusters of spins at oxygen vacancies, for example, would be expected to be superparamagnetic.

We now consider some possible explanations, illustrated schematically in Figure 9. All are consistent with a surface origin of the magnetism, and its association with oxygen vacancies, as demonstrated by the vacuum, oxygen and Tiron treatments. The loss of oxygen from the surface on heating in vacuum can be written as:

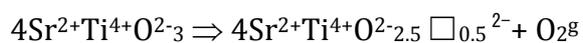

$4Sr^{2+}Ti^{4+}O^{2-}{}_3 \Rightarrow 4Sr^{2+}Ti^{4+}O^{2-}{}_{2.5} \square_{0.5}{}^{2-} + O_2{}^g$

where the oxygen vacancies □ formed in the perovskite lattice during vacuum treatment are negatively charged because the departing oxygen gas is uncharged. Their density may be lower than suggested by the brownmillerite formula, but it can hardly be higher. The electrons associated with the oxygen vacancy are thought to be localized onto the adjacent titanium sites, or delocalized into a broad $d_{xy}$ band, at least at (100) surfaces [10-12]. In the reaction with Tiron, the two OH groups attached to the phenol ring tend to bind to surface titanium ions, completing their oxygen coordination octahedron by filling oxygen vacancies at the surface, thereby eliminating the source of the magnetism. Schematically,

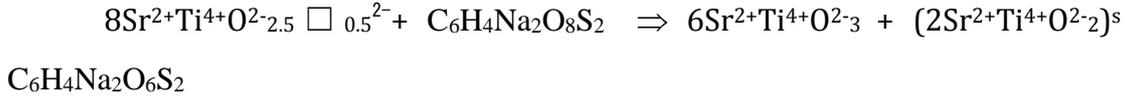

$8Sr^{2+}Ti^{4+}O^{2-}{}_{2.5} \square_{0.5}{}^{2-} + C_6H_4Na_2O_8S_2 \Rightarrow 6Sr^{2+}Ti^{4+}O^{2-}{}_3 + (2Sr^{2+}Ti^{4+}O^{2-}{}_2)^s C_6H_4Na_2O_6S_2$

The models we consider are:
a) Conventional Heisenberg ferromagnetism due to localized electrons with $s = ½$ situated near oxygen defects.
b) Zener ferromagnetism of $2p$ oxygen holes.
c) Stoner ferromagnetism of electrons in a defect-related impurity band.
d) Giant spin magnetism associated with a noncollinear spin texture.
e) Giant orbital paramagnetism associated with surface currents.

a) Localized spin moments have been found to form near oxygen vacancies in many calculations [5-8, 10-12, 54, 55] Pure $s$-electrons have no orbital moment and should not exhibit hysteresis, but the localized electrons acquire some Ti($3d$) character by hybridization with adjacent empty Ti($e_g$) orbitals. To explain our observations, we would need the localized spins to be present in high concentrations near the surface, with sufficiently strong ferromagnetic exchange coupling to produce a Curie temperature of order 1000 K. This is a tall order. The maximum possible oxygen defect concentration in SrTiO$_3$ is $x_v = 1/6$, corresponding to the brownmillerite structure SrTiO$_{2.5}$[4]. The necessary exchange $J$ may be estimated from the standard mean-field expression

$$T_c = 2x_v Z J s(s + 1)/3k_B \qquad (1)$$

where $Z$ is the coordination number. An immediate problem is that $x_v = 1/6$ is far below the percolation threshold $x_1 = 0.31$ for nearest neighbours in the simple-cubic oxygen lattice, and even below the threshold $x_2 = 0.20$ if second-neighbour sites are taken into account. Assuming $Z = 12$, $J$ should be ~1000 K. This is an unprecedentedly large positive value, and opposite in sign to the anticipated antiferromagnetic superexchange. The model also requires three layers of the whole SrTiO$_3$ surface to have brownmillerite stoichiometry to provide a moment of 1.5 $\mu_B$ per surface unit cell. , which is not observed [56]. Computed values of the magnetic ordering temperature of ferromagnetic clusters of electrons trapped at oxygen vacancies are no more than 50 K [55].

**b**) Another possibility is that ferromagnetism could be associated with holes at the top of the 2$p$(O) band, which are associated with 2$p^5$ O$^-$ ions at the surface. Electron hopping with spin memory among the ligand hole states could align the moments of the 2$p^5$ oxygen anions by a process analogous to Zener double exchange. While it has been proposed that cation vacancies can lead to such ligand hole states [57], the defects expected at reduced SrTiO$_3$ surfaces, or in the bulk, are oxygen vacancies, not cation vacancies. Although the oxygen molecule itself has a triplet ground state, superoxides such as CsO$_2$ where the magnetism is associated with peroxide ions are antiferromagnets, with Néel temperatures of order 10 K.

**c)** A more plausible idea is to associate high-temperature surface ferromagnetism with the oxygen-vacancy-related impurity band that forms just below the Ti($t_{2g}$) conduction band. This avoids the problems of percolation and unrealistic surface stoichiometry. If it is narrow and spin-split, but not exactly half full, the impurity band could be half-metallic, and largely immune to spin-wave excitations [58]. The Curie temperature in the Stoner model can then be very high, of order the bandwidth, which could easily be 0.1 eV or more. The impurity band might have no orbital moment, and hence no magnetocrystalline anisotropy to lead to coercivity, although this should not be the case if the band has $t_{2g}$ orbital character. However, in order to explain the temperature-independent magnetization, any coupling with the $s$ = ½ localized spins in the oxygen vacancies should either be very strong and ferromagnetic, or else vanishingly small (if it was antiferromagnetic there would be no net moment). Neither seems reasonable, but if the localized and itinerant moments were weakly coupled, there should be an additional paramagnetic moment saturating at low temperatures in high fields at a value equal to the ferromagnetic moment. This is not seen (Fig. 2c). If pairs of midgap states on neighbouring titanium ions [12] formed spin dimers, the ferromagnetic moment would be due to a spin-split $d_{xy}$ band in a surface layer.

**d)** A recent idea is a combination of a chiral spin texture with ferromagnetically-coupled local moments as suggested by the SARPES data [20, 39], and some recent spin-polarized electronic structure calculations [10, 11]. The magnetization process would then amount to deforming a long period real-space spin texture in an applied magnetic field of order 1 T. The DFT calculations are for small blocks of unit cells and cannot capture a long-period structure. The calculated spin moments are generally smaller than we observe, and they are located in the surface layer, giving a large surface magnetization that is not manifest in the dependence of the magnetization curve on the orientation of the applied field. Furthermore, the thermal stability of these electronic spin structures induced by exchange and spin-obit interactions among single electrons with partial 3$d$ character is uncertain; there are estimates of a stabilization energy of the magnetic state of 80 – 160 K per formula unit for low oxygen vacancy concentrations of about 1:50 for a divacancy and a single vacancy, respectively [11]. The possibility of a long-period modulated spin texture having a very high magnetic ordering temperature might account for anhysteretic saturating magnetization curves, but it would be unrelated to the Rashba spin-orbit coupling, which is strongly anisotropic.

**e)** The last explanation we consider is completely different. It is that the temperature-independent 'ferromagnetic' response to an applied field is unrelated to any sort of collective, exchange driven ferromagnetic spin ordering, but is entirely field-induced and originates from coherent mesoscopic conduction currents. In other words, we are seeing an unusual type of temperature-independent, saturating orbital paramagnetism. This possibility has been evoked in the context of Au nanoparticles [59], and the magnetic consequences of the theory [60] have recently been developed in the context of $CeO_2$ nanoparticles [61]. There the magnetization curves are fitted to the theoretical expression

$$M = M_s\, x(1 + x^2)^{-1/2} \qquad (2)$$

where $x = \mathcal{C}B$. The theory is based on the formation of coherent states of large numbers of spinless electrons in response to zero-point fluctuations of the vacuum electromagnetic field. In the present case, fitting yields values of $\mathcal{C}$ in the range 3 - 5 T$^{-1}$. The characteristic wavelength given by the theory is

$$\lambda = [(\mathcal{C}/M_s)(6\hbar c\, f_c)]^{1/4} \qquad (3)$$

where $f_c$ is taken as the volume fraction of the sample that is magnetically coherent. A first estimate of $f_c$ is the volume where the oxygen vacancies are located; assuming a defective layer ~ 2 nm thick at the surface, a surface moment of 5 $\mu_B$/nm$^2$ corresponds to a magnetization of the surface region of approximately 28 kAm$^{-1}$. Thus with $\mathcal{C}$ = 4 T$^{-1}$ we we find λ ≈ 72 nm (17 eV). If, however, we determine the thickness selfconsistently by assuming that the thickness of the magnetic region is equal to $\lambda$, $f_c \approx 2\lambda/t$ where $t$ is the substrate thickness (0.5 mm), we find

$$\lambda = [(C/M_s)(12\hbar c/t)]^{1/3} \qquad (4)$$

or $\lambda$ = 111 nm. The corresponding energy is 11.2 eV. The interpretation of $\lambda$, according to the theory, which is only applicable to quasi-two-dimensional systems like reduced SrTiO$_3$ with a thin active surface area, is that it is the size of coherent electronic domains formed by interaction with zero-point fluctuations of the vacuum electromagnetic field [60]. An energy of 12 eV corresponds to a charge-transfer excitation that has been observed in SrTiO$_3$ crystals by inelastic X-ray scattering[62].

It is a problem explain why, in *any* model with a thin spontaneous or induced magnetic surface layer having a moment of up to 1.5 $\mu_B$ per surface unit cell, the data fail to show the anticipated anisotropy of the slope of the initial magnetization curves of Fig 8b) and 8c). These magnetization curves appear to be strictly isotropic. The magnetization depends on the moment per unit area and the thickness of the surface magnetic layer, which creates a demagnetizing field that depends on the orientation of the magnetization relative to the surface. A surface layer just two unit cells thick has a magnetization of 120 kAm$^{-1}$, and there should be a difference of this magnitude in the saturation fields in the parallel and perpendicular magnetization curves. Ideally, the parallel magnetization curve should saturate quickly in a low field, and the perpendicular curve should require a field equal to the magnetization for saturation, as is usually found in ferromagnetic thin films. 110 crystals from supplier 'A' and crystals of all three orientations from supplier 'C' have been measured, but there are no perceptible differences between the parallel and perpendicular curves. The extrapolated saturation exceeds the ferromagnetic saturation magnetization, so It follows that the magnetic order in the spin exchange models a) – d) cannot not be strictly ferromagnetic. In model e), the induced magnetism is spread over a thicker surface layer and the magnetization is correspondingly reduced is able to fit the data, but it is a problem to understand how conduction could persist to depths of order 100 nm at below STO surface.

We have emphasised the essentially anhysteretic nature of the temperature-independent magnetic signal we are seeking to explain. However, in a few samples we have observed temperature-independent hysteresis with a remanence ratio of about 20%, which suggests that the Rashba spin-orbit interaction could couple spin moments to the field-induced orbital moments, and lead to the hysteresis that we occasionally observe.

A summary of strengths and weaknesses of the various models for the high-temperature magnetism is given in Table 3.

Table 3. Models for the magnetism associated with $SrTiO_3$ surfaces.

| Model | Strengths | Weaknesses |
| --- | --- | --- |
| Heisenberg ferromagnetism of $s = ½$ localized electrons | $S = ½$ midgap states are predicted by DFT calculations | Exchange is usually antiferro-magnetic. Predicted $T_c$ is much too low. Percolation problem |
| Zener ferromagnetism of $O^-$ ions | Strong correlations in $2p$ shell might give high $T_C$ | $2p$ band is full. No evidence for high $T_C$ in peroxides. |
| Stoner ferromagnetism of a defect-related impurity band | High $T_C$ is possible. Possibly no anisotropy | Paramagnetism of associated localized electrons is not seen |
| Ferrochiral spin texture related to Rashba spin-orbit coupling | The necessary localized and itinerant electrons are provided by oxygen vacancies. | Thermal stability at $T \sim 1000$ K is unknown. Huge Rashba splitting is unconfirmed, and has to be anisotropic. |
| Giant orbital paramagnetism | Inherently temperature-independent & anhysteretic. Predicts magnetization curves  No spontaneous magnetism | Conduction required in a relatively thick surface layer to account for isotropic magnetization curves. |

**4. Conclusions**

Our study illustrates how much information can be gleaned from a critical analysis of magnetization curves. The main conclusions are the following:

1) It is important to take care when sourcing substrates if small magnetic signals associated with thin film heterostructures grown on $SrTiO_3$ crystals are of interest. Most cuts contain extraneous ferromagnetic impurities in the surface at the ppm level, which may have a magnetic moment comparable to that of the thin films themselves. From a magnetic viewpoint, we have found no cut that is consistently clean regardless of source. It is important to characterize the substrates separately when measuring small magnetic signals, and to always run blanks. Also be aware that low-pressure, high-temperature deposition processes may induce an extra temperature independent 'ferromagnetic' signal due to the STO substrate. This warning is specific to $SrTiO_3$. Other oxide substrates such as sapphire are much less problematic [48].

2) The 'ferromagnetic' surface signal is an unprecedented intrinsic high-temperature (or temperature-independent) magnetic phenomenon. It is clearly associated with oxygen vacancies in a surface layer, and it is sensitive to the surface electron concentration. These quantities can be altered by gating [63, 64], or as we have shown, by treatment with the catchecol electron donor molecule Tiron.

3) It is unlikely that to that the high-temperature intrinsic magnetic signal at $SrTiO_3$ surfaces is solely associated with localized spins trapped at oxygen vacancies, or with ligand holes. It is more plausible to associate it with a half-metallic defect-related impurity band, provided the moments of any associated localized electrons trapped at the oxygen vacancies are silent. A hybrid explanation in terms of a noncollinear spin texture produced by ferromagnetic coupling of localized, oxygen-vacancy-related mid-gap spin states and a chiral Rashba texture of delocalized $d_{xy}$ electron spins in surface quantum well states would require the spin structure to be thermally stable up temperatures of order 1000 K, which has yet to be demonstrated, and a strongly anisotropic magnetization, which is not observed in the crystals. The lack of evidence of anisotropic demagnetizing fields associated with thin ferromagnetic surface layers is unexplained in any case.

4) A completely different explanation in terms of giant orbital paramagnetism arising from coherent orbital currents is able to account for the magnetization data, provided the thickness of the magnetic surface layer exceeds about 10 nm. Both this orbital model, and the spin texture model depend critically on a mobile surface electron concentration.

*Methods*

The crystals were handled using only plastic or wood. Powders were produced in an agate mortar used only for the purpose. The annealing of crystals and powders was done in alumina boats in a clean fused silica tube under a vacuum of $10^{-5}$ Pa produced by a turbomolecular pump. Crystals of mass ~40 – 80 mg were mounted directly in long plastic straws, which make no measurable contribution to the measured signal. The powders were tightly confined in the hemispherical space between two interlocking gelcaps of mass 30 ± 1 mg, which were then mounted in a plastic straw. The magnetic signal from the gelcaps is perfectly diamagnetic, with slope -200 $10^{-15}$ m$^3$ (equivalent to –145 $10^{-9}$ Am$^2$T$^{-1}$, or –14.5 μemu/kOe) which is several times the diamagnetism of the SrTiO$_3$ powder. All the measurements were made in an MPMS XL5 Quantum Design SQUID magnetometer. MWH135 Crystal-Clear wax (QuickStick 135) wax (South Bay Technology, San Clemente, California) was used to mount the crystals for lapping.

*Acknowledgements.*

We are grateful to Lorena Monzon for suggesting the use of Tiron., and to Stephen Porter and Pelin Tozman for some of the magnetic measurements. Cora McKenna of TCD Geochemistry carried out the laser microprobe mass spectroscopy, and Lorna Eades of Edinburgh University, School of Chemistry kindly did the ICPMS. We are grateful to Ariando, Z Q Liu and T Venkatesan for helpful discussions.

**Figure Captions.**

Figure 1. Materials used in this study: a) The perovskite structure of SrTiO$_3$; b) the schematic electronic structure; c) single crystals, as-received and reduced in vacuum; d) powders made from the crystals before and after vacuum reduction; e) Electron micrograph of a powder.

Figure 2. Some representative magnetization signals: a) As measured magnetization of sample 100 (A), as received crystal, crystal after vacuum annealing, powder and powder after vacuum annealing. b) Data corrected for the background diamagnetism of SrTiO$_3$ and the gelcap used to hold the powder. c) Temperature dependent, hysteretic magnetization curves of a 110 [A] crystal. d) Temperature dependence of magnetization of 100 [A] annealed crystal powders and e) Temperature-dependent susceptibility of 100 crystals from four different suppliers.

Figure 3. Summary of specific magnetization for 100, 110 and 111 crystals from four different suppliers: Blue and red bars are for samples before and after heating in vacuum at 750° C for 2 hrs. Solid bars are for single crystals, speckled bars are for corresponding powders. Note that the data are plotted on a log scale. The light dashed lines correspond to the signal levels expected for 1 ppm and 10 ppm of iron metal impurity. In fact, all iron contents for 100 crystals are < 0.5 ppm, except for Supplier 'B', where it is 5 ppm.

Figure 4. Dependences of crystal magnetic moments on annealing in vacuum: Data are for 100, 110 and 111 crystals from supplier A. The largest magnetic moment is found when 110 or 111 crystals are vacuum-annealed at 600 – 750 °C. No moment appears at any temperature for 100 crystals

Figure 5. Surface chemical analysis: Concentrations scans of Fe, Co and Ni measured across the surface of a 100 [A] crystal by laser ablation mass spectroscopy. The distribution is very non-uniform, suggesting that particles of a transition-meta-rich impurity phase are embedded in the surface

Figure 6. Variation of the magnetic moment of 110[A] crystals annealed at 750°C after surface treatment. a) mechanical thinning from 0.5 mm to 0.15 mm b) hot

wax treatment at 150°C on one or both sides c) treatment in molten NaCl at 900°C with the effect of reannealing at 750°C.

Figure 7. Treatments of annealed SrTiO$_3$ single-crystal powders: Effects of a) annealing 111 (C) powder in vacuum and reannealing the reduced powder in oxygen, b) immersion of the reduced powder in ethylene glycol and TIRON

Figure 8 Magnetization curves of SrTiO$_3$ crystals as a function of applied field direction: a) 110 [A]; b) 110 [C]; c) 111 [A]

Figure 9. Schematic depiction of models considered to explain the high-temperature anysteretic magnetic signal originating from the SrTiO$_3$ surface: a) Heisenberg ferromagnetism of electrons trapped at oxygen vacancies; b) Zener ferromagnetism of O$^-$ ions due to hole hopping in a 2p band c) Stoner ferromagnetism of a spin-split defect-related impurity band  d) giant ferrochiral spin texture and e) Giant orbital paramagnetism. The five models are discussed in the text, and summarized in Table 3.

[1] J.A. Enterkin, A.K. Subramanian, B.C. Russell, M.R. Castell, K.R. Poeppelmeier, L.D. Marks, A homologous series of structures on the surface of SrTiO$_3$ (110), Nature materials, 9 (2010) 245-248.
[2] Z. Wang, Z. Zhong, X. Hao, S. Gerhold, B. Stöger, M. Schmid, J. Sánchez-Barriga, A. Varykhalov, C. Franchini, K. Held, U. Diebold, Anisotropic two-dimensional electron gas at SrTiO$_3$(110), Proceedings of the National Academy of Sciences, 111 (2014) 3933-3937.
[3] A.E. Becerra-Toledo, J.A. Enterkin, D.M. Kienzle, L.D. Marks, Water adsorption on SrTiO$_3$(001): II. Water, water, everywhere, Surface Science, 606 (2012) 791-802.
[4] M.A.A. Franco, M.V. Regi, Anion deficiency in strontium titanate, Nature, 270 (1977) 706-708.
[5] C. Lin, A.A. Demkov, Electron Correlation in Oxygen Vacancy in SrTiO$_3$, Physical Review Letters, 111 (2013) 217601.
[6] A. Lopez-Bezanilla, P. Ganesh, P.B. Littlewood, Research Update: Plentiful magnetic moments in oxygen deficient SrTiO$_3$, APL Materials, 3 (2015) 100701.
[7] A. Lopez-Bezanilla, P. Ganesh, P.B. Littlewood, Magnetism and metal-insulator transition in oxygen-deficient SrTiO$_3$, Physical Review B, 92 (2015) 115112.
[8] Z. Hou, K. Terakura, Defect states induced by oxygen vacancies in cubic SrTiO$_3$: first-principles calculations, Journal of the Physical Society of Japan, 79 (2010) 114704.
[9] X. Hao, Z. Wang, M. Schmid, U. Diebold, C. Franchini, Coexistence of trapped and free excess electrons in SrTiO$_3$, Physical Review B, 91 (2015) 085204.
[10] A.C. Garcia-Castro, M.G. Vergniory, E. Bousquet, A.H. Romero, Spin texture induced by oxygen vacancies in strontium perovskite (001) surfaces: A theoretical comparison between SrTiO$_3$ and SrHfO$_3$, Physical Review B, 93 (2016) 045405.
[11] M. Altmeyer, H.O. Jeschke, O. Hijano-Cubelos, C. Martins, F. Lechermann, K. Koepernik, A.F. Santander-Syro, M.J. Rozenberg, R. Valentí, M. Gabay, Magnetism, Spin Texture, and In-Gap States: Atomic Specialization at the Surface of Oxygen-Deficient SrTiO$_3$, Physical Review Letters, 116 (2016) 157203.
[12] H.O. Jeschke, J. Shen, R. Valentí, Localized versus itinerant states created by multiple oxygen vacancies in SrTiO$_3$, New Journal of Physics, 17 (2015) 023034.
[13] A. Ohtomo, H.Y. Hwang, A high-mobility electron gas at the LaAlO$_3$/SrTiO$_3$ heterointerface, Nature, 427 (2004) 423-426.
[14] N. Nakagawa, H.Y. Hwang, D.A. Muller, Why some interfaces cannot be sharp, Nature materials, 5 (2006) 204-209.
[15] G. Herranz, M. Basletic, M. Bibes, C. Carretero, E. Tafra, E. Jacquet, K. Bouzehouane, C. Deranlot, A. Hamzic, J.M. Broto, A. Barthelemy, A. Fert, High mobility in LaAlO$_3$/SrTiO$_3$ heterostructures: origin, dimensionality, and perspectives, Physical Review Letters, 98 (2007) 216803.
[16] Z.Q. Liu, C.J. Li, W.M. Lü, X.H. Huang, Z. Huang, S.W. Zeng, X.P. Qiu, L.S. Huang, A. Annadi, J.S. Chen, J.M.D. Coey, T. Venkatesan, Ariando, Origin of the Two-Dimensional Electron Gas at LaAlO$_3$/SrTiO$_3$ Interfaces: The Role of Oxygen Vacancies and Electronic Reconstruction, Physical Review X, 3 (2013) 021010.
[17] W. Meevasana, P.D.C. King, R.H. He, S.K. Mo, M. Hashimoto, A. Tamai, P. Songsiriritthigul, F. Baumberger, Z.X. Shen, Creation and control of a two-dimensional electron liquid at the bare SrTiO$_3$ surface, Nature Materials, 10 (2011) 114-118.
[18] A.F. Santander-Syro, O. Copie, T. Kondo, F. Fortuna, S. Pailhes, R. Weht, X.G. Qiu, F. Bertran, A. Nicolaou, A. Taleb-Ibrahimi, P. Le Fevre, G. Herranz, M. Bibes, N. Reyren, Y. Apertet, P. Lecoeur, A. Barthelemy, M.J. Rozenberg, Two-

dimensional electron gas with universal subbands at the surface of SrTiO$_3$, Nature, 469 (2011) 189-193.

[19] N.C. Plumb, M. Salluzzo, E. Razzoli, M. Månsson, M. Falub, J. Krempasky, C.E. Matt, J. Chang, M. Schulte, J. Braun, H. Ebert, J. Minár, B. Delley, K.J. Zhou, T. Schmitt, M. Shi, J. Mesot, L. Patthey, M. Radović, Mixed Dimensionality of Confined Conducting Electrons in the Surface Region of SrTiO$_3$, Physical Review Letters, 113 (2014) 086801.

[20] A.F. Santander-Syro, F. Fortuna, C. Bareille, T.C. Rödel, G. Landolt, N.C. Plumb, J.H. Dil, M. Radović, Giant spin splitting of the two-dimensional electron gas at the surface of SrTiO$_3$, Nature Materials, 13 (2014) 1085-1090.

[21] Z. Zhong, A. Tóth, K. Held, Theory of spin-orbit coupling at LaAlO3/SrTiO3 interfaces and SrTiO$_3$ surfaces, Physical Review B, 87 (2013) 161102.

[22] P.D.C. King, S. McKeown Walker, A. Tamai, A. de la Torre, T. Eknapakul, P. Buaphet, S.K. Mo, W. Meevasana, M.S. Bahramy, F. Baumberger, Quasiparticle dynamics and spin–orbital texture of the SrTiO$_3$ two-dimensional electron gas, Nature Communications, 5 (2014) 3414.

[23] G. Khalsa, B. Lee, A.H. MacDonald, Theory of t$_{2g}$ electron-gas Rashba interactions, Physical Review B, 88 (2013) 041302.

[24] A. Brinkman, M. Huijben, M. van Zalk, J. Huijben, U. Zeitler, J.C. Maan, W.G. van der Wiel, G. Rijnders, D.H. Blank, H. Hilgenkamp, Magnetic effects at the interface between non-magnetic oxides, Nature materials, 6 (2007) 493-496.

[25] M. Lee, J.R. Williams, S. Zhang, C.D. Frisbie, D. Goldhaber-Gordon, Electrolyte gate-controlled Kondo effect in SrTiO$_3$, Physical Review Letters, 107 (2011) 256601.

[26] Y. Lee, C. Clement, J. Hellerstedt, J. Kinney, L. Kinnischtzke, X. Leng, S.D. Snyder, A.M. Goldman, Phase diagram of electrostatically doped SrTiO$_3$, Physical Review Letters, 106 (2011) 136809.

[27] M. Li, T. Graf, T.D. Schladt, X. Jiang, S.S. Parkin, Role of percolation in the conductance of electrolyte-gated SrTiO$_3$, Phys Rev Lett, 109 (2012) 196803.

[28] P. Moetakef, J.R. Williams, D.G. Ouellette, A.P. Kajdos, D. Goldhaber-Gordon, S.J. Allen, S. Stemmer, Carrier-Controlled Ferromagnetism in SrTiO$_3$, Physical Review X, 2 (2012).

[29] B. Kalisky, J.A. Bert, B.B. Klopfer, C. Bell, H.K. Sato, M. Hosoda, Y. Hikita, H.Y. Hwang, K.A. Moler, Critical thickness for ferromagnetism in LaAlO$_3$/SrTiO$_3$ heterostructures, Nature Commununications, 3 (2012) 922.

[30] J.A. Bert, B. Kalisky, C. Bell, M. Kim, Y. Hikita, H.Y. Hwang, K.A. Moler, Direct imaging of the coexistence of ferromagnetism and superconductivity at the LaAlO$_3$/SrTiO$_3$ interface, Nature Physics, 7 (2011) 767-771.

[31] N. Pavlenko, T. Kopp, E.Y. Tsymbal, J. Mannhart, G.A. Sawatzky, Oxygen vacancies at titanate interfaces: Two-dimensional magnetism and orbital reconstruction, Physical Review B, 86 (2012) 064431.

[32] N. Pavlenko, T. Kopp, E.Y. Tsymbal, G.A. Sawatzky, J. Mannhart, Magnetic and superconducting phases at the LaAlO$_3$/SrTiO$_3$ interface: The role of interfacial Ti 3*d* electrons, Physical Review B, 85 (2012) 020407.

[33] M. Salluzzo, S. Gariglio, D. Stornaiuolo, V. Sessi, S. Rusponi, C. Piamonteze, G.M. De Luca, M. Minola, D. Marre, A. Gadaleta, H. Brune, F. Nolting, N.B. Brookes, G. Ghiringhelli, Origin of interface magnetism in BiMnO$_3$/SrTiO$_3$ and LaAlO$_3$/SrTiO$_3$ heterostructures, Physical Review Letters, 111 (2013) 087204.

[34] Ariando, X. Wang, G. Baskaran, Z.Q. Liu, J. Huijben, J.B. Yi, A. Annadi, A.R. Barman, A. Rusydi, S. Dhar, Y.P. Feng, J. Ding, H. Hilgenkamp, T. Venkatesan, Electronic phase separation at the LaAlO$_3$/SrTiO$_3$ interface, Nature Communications, 2 (2011) 188.

[35] Z.Q. Liu, W.M. Lü, S.L. Lim, X.P. Qiu, N.N. Bao, M. Motapothula, J.B. Yi, M. Yang, S. Dhar, T. Venkatesan, Ariando, Reversible room-temperature ferromagnetism in Nb-doped SrTiO$_3$ single crystals, Physical Review B, 87 (2013) 220405.

[36] H. Trabelsi, M. Bejar, E. Dhahri, M. Sajieddine, M.A. Valente, A. Zaoui, Effect of the oxygen deficiencies creation on the suppression of the diamagnetic behavior of SrTiO$_3$ compound, Journal of Alloys and Compounds, 680 (2016) 560-564.

[37] S.S. Rao, Y.F. Lee, J.T. Prater, A.I. Smirnov, J. Narayan, Laser annealing induced ferromagnetism in SrTiO$_3$ single crystal, Applied Physics Letters, 105 (2014) 042403.

[38] F. Bi, M. Huang, S. Ryu, H. Lee, C.-W. Bark, C.-B. Eom, P. Irvin, J. Levy, Room-temperature electronically-controlled ferromagnetism at the LaAlO$_3$/SrTiO$_3$ interface, Nature Communications, 5 (2014) 5019.

[39] S.M. Walker, S. Ricco, F.Y. Bruno, A.d.l. Torre, A. Tamai, E. Golias, A. Varykhalov, M.H. D. Marchenko, M.S. Bahramy, P.D.C. King, J. Sanchez-Barriga, and F. Baumberger, Absence of Giant Spin Splitting in the Two-Dimensional Electron Liquid at the Surface of SrTiO$_3$ (001), arXiv:1603.00181v1 [cond-mat.str-el], (2016).

[40] L.P. Gor'kov, Antiferromagnetism of two-dimensional electronic gas on light-irradiated SrTiO$_3$ and at LaAlO$_3$/SrTiO$_3$ interfaces, Journal of Physics: Condensed Matter, 27 (2015) 252001.

[41] S.M.M. Yee, D.A. Crandles, L.V. Goncharova, Ferromagnetism on the unpolished surfaces of single crystal metal oxide substrates, Journal of Applied Physics, 110 (2011) 033906.

[42] M.A. Garcia, E. Fernandez Pinel, J. de la Venta, A. Quesada, V. Bouzas, J.F. Fernández, J.J. Romero, M.S. Martín González, J.L. Costa-Krämer, Sources of experimental errors in the observation of nanoscale magnetism, Journal of Applied Physics, 105 (2009) 013925.

[43] D.W. Abraham, M.M. Frank, S. Guha, Absence of magnetism in hafnium oxide films, Applied Physics Letters, 87 (2005) 252502.

[44] Y. Belghazi, G. Schmerber, S. Colis, J.L. Rehspringer, A. Dinia, A. Berrada, Extrinsic origin of ferromagnetism in ZnO and Zn$_{0.9}$Co$_{0.1}$O magnetic semiconductor films prepared by sol-gel technique, Applied Physics Letters, 89 (2006) 122504.

[45] J.M.D. Coey, Dilute magnetic oxides, Current Opinion in Solid State and Materials Science, 10 (2006) 83-92.

[46] K. Potzger, S. Zhou, Non-DMS related ferromagnetism in transition metal doped zinc oxide, Physica Status Solidi B, 246 (2009) 1147-1167.

[47] M. Khalid, A. Setzer, M. Ziese, P. Esquinazi, D. Spemann, A. Pöppl, E. Goering, Ubiquity of ferromagnetic signals in common diamagnetic oxide crystals, Physical Review B, 81 (2010) 214414.

[48] M. Venkatesan, P. Kavle, S.B. Porter, K. Ackland, J.M.D. Coey, Magnetic Analysis of Polar and Nonpolar Oxide Substrates, IEEE Transactions on Magnetics, 50 (2014) 2201704.

[49] M.S. Ata, Y. Liu, I. Zhitomirsky, A review of new methods of surface chemical modification, dispersion and electrophoretic deposition of metal oxide particles, RSC Advances, 4 (2014) 22716.

[50] J.M.D. Coey, Magnetism and Magnetic Materials, Cambridge University Press p.75, Cambridge, 2010.

[51] H.P.R. Frederikse, G.A. Candela, Magnetic Susceptibility of Insulating and Semiconducting Strontium Titanate, Physical Review, 147 (1966) 583-584.

[52] J.M.D. Coey, J.T. Mlack, M. Venkatesan, P. Stamenov, Magnetization Process in Dilute Magnetic Oxides, IEEE Transactions on Magnetics, 46 (2010) 2501.
[53] W.D. Rice, P. Ambwani, M. Bombeck, J.D. Thompson, G. Haugstad, C. Leighton, S.A. Crooker, Persistent optically induced magnetism in oxygen-deficient strontium titanate, Nature Materials, 13 (2014) 481-487.
[54] H. Choi, J.D. Song, K.-R. Lee, S. Kim, Correlated Visible-Light Absorption and Intrinsic Magnetism of $SrTiO_3$ Due to Oxygen Deficiency: Bulk or Surface Effect?, Inorganic Chemistry, 54 (2015) 3759-3765.
[55] C. Lin, A.A. Demkov, Consequences of oxygen-vacancy correlations at the $SrTiO_3$ interface, Physical Review Letters, 113 (2014) 157602.
[56] N. Erdman, K.R. Poeppelmeier, M. Asta, O. Warschkow, D.E. Ellis, L.D. Marks, The Structure and Chemistry of the $TiO_2$-Rich Surface of $SrTiO_3$(001). Nature, 419 (2002) 55-58.
[57] I.S. Elfimov, S. Yunoki, G.A. Sawatzky, Possible Path to a New Class of Ferromagnetic and Half-Metallic Ferromagnetic Materials, Physical Review Letters, 89 (2002) 216403.
[58] D.M. Edwards, M.I. Katsnelson, High-temperature ferromagnetism of sp electrons in narrow impurity bands: application to $CaB_6$, Journal of Physics: Condensed Matter, 18 (2006) 7209-7225.
[59] R. Gréget, G.L. Nealon, B. Vileno, P. Turek, C. Mény, F. Ott, A. Derory, E. Voirin, E. Rivière, A. Rogalev, F. Wilhelm, L. Joly, W. Knafo, G. Ballon, E. Terazzi, J.-P. Kappler, B. Donnio, J.-L. Gallani, Magnetic Properties of Gold Nanoparticles: A Room-Temperature Quantum Effect, ChemPhysChem, 13 (2012) 3092-3097.
[60] S. Sen, K.S. Gupta, J.M.D. Coey, Mesoscopic structure formation in condensed matter due to vacuum fluctuations, Physical Review B, 92 (2015) 155115.
[61] J.M.D. Coey, K. Ackland, M. Venkatesan, S. Sen, Collective magnetic response of $CeO_2$ nanoparticles, Nature Physics, (2016).
[62] S. Kawakami, N. Nakajima, T. Takigawa, M. Nakatake, H. Maruyama, Y. Tezuka, T. Iwazumi, UV-Induced Change in the Electronic Structure of $SrTiO_3$ at Low Temperature Probed by Resonant X-ray Emission Spectroscopy, Journal of the Physical Society of Japan, 82 (2013) 1-4.
[63] M. Li, W. Han, X. Jiang, J. Jeong, M.G. Samant, S.S.P. Parkin, Suppression of Ionic Liquid Gate-Induced Metallization of $SrTiO_3$(001) by Oxygen, Nano Letters, 13 (2013) 4675.
[64] S. Shimizu, S. Ono, T. Hatano, Y. Iwasa, Y. Tokura, Enhanced cryogenic thermopower in $SrTiO_3$ by ionic gating, Physical Review B, 92 (2015) 165304.

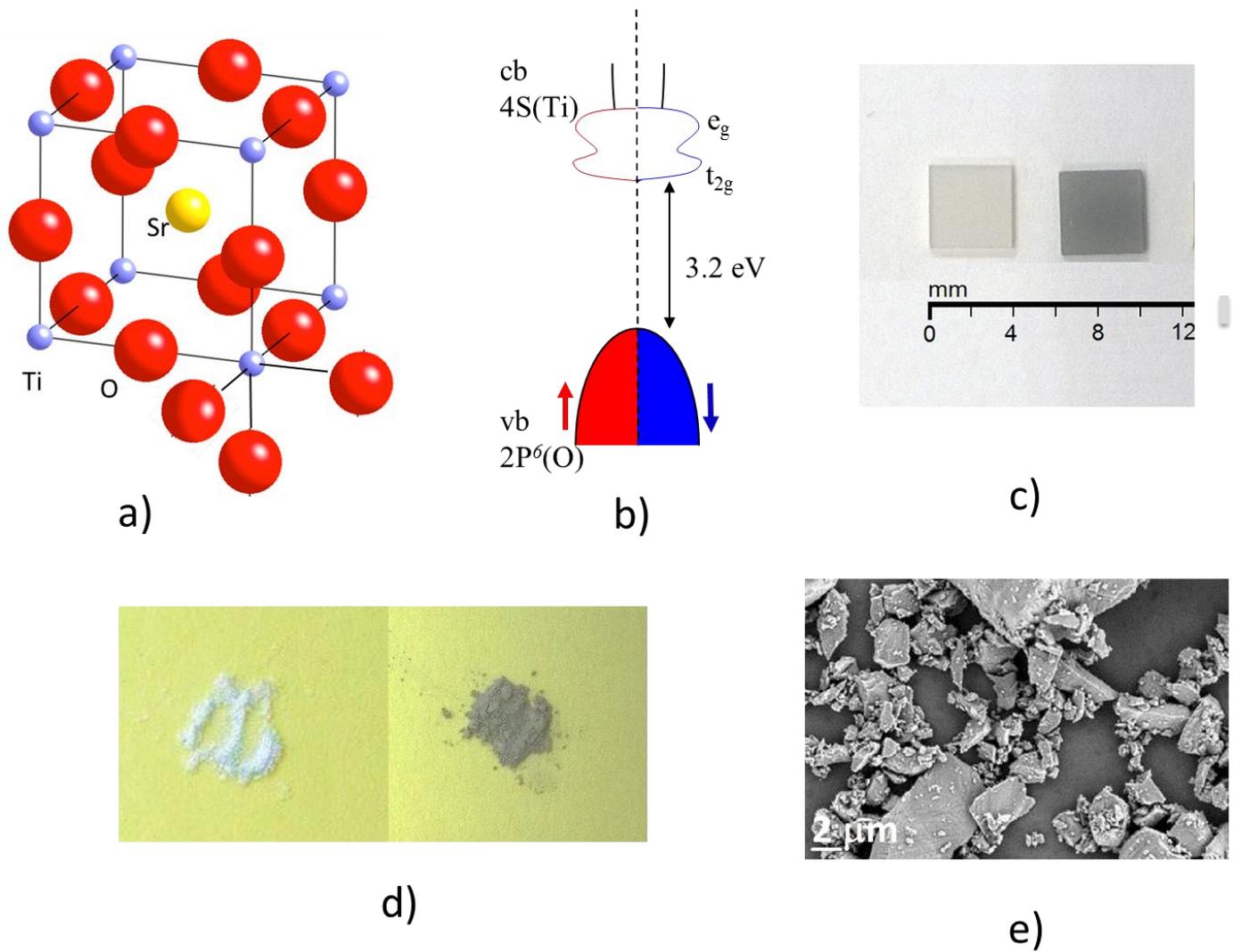

Figure 1. Materials used in this study: a) The perovskite structure of $SrTiO_3$; b) the schematic electronic structure; c) single crystals, as-received and reduced in vacuum; d) powders made from the crystals before and after vacuum reduction; e) Electron micrograph of a powder.

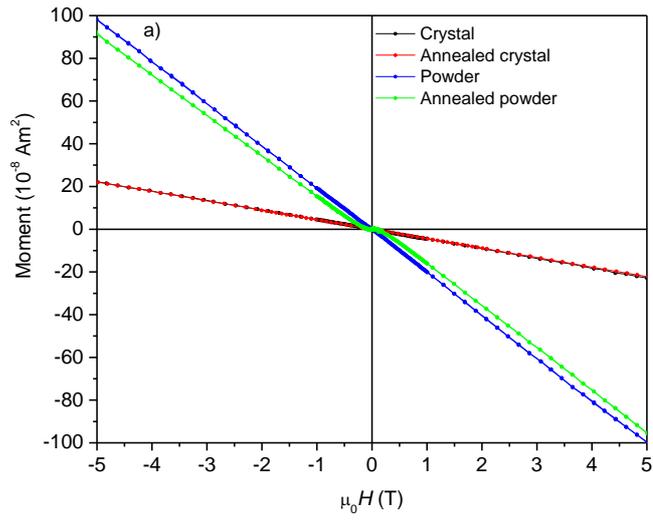
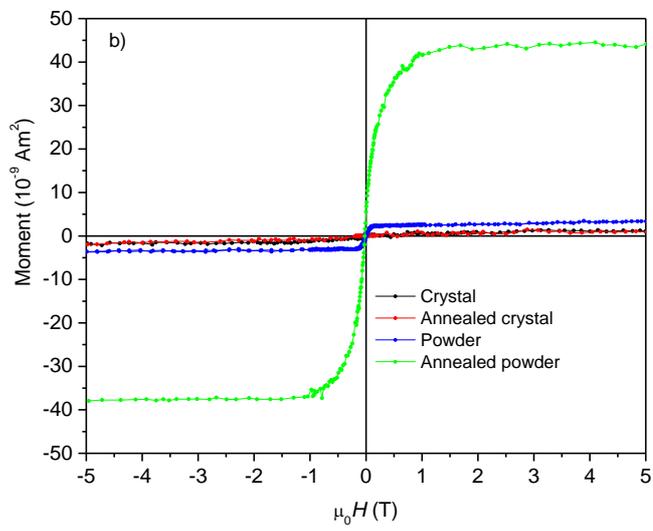
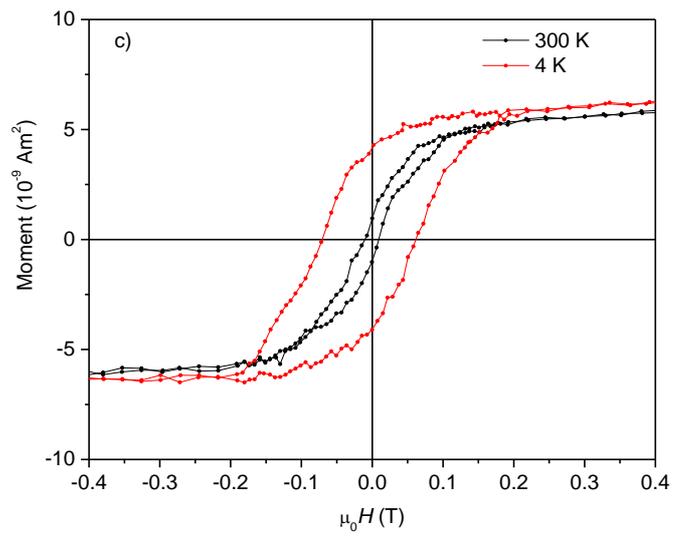

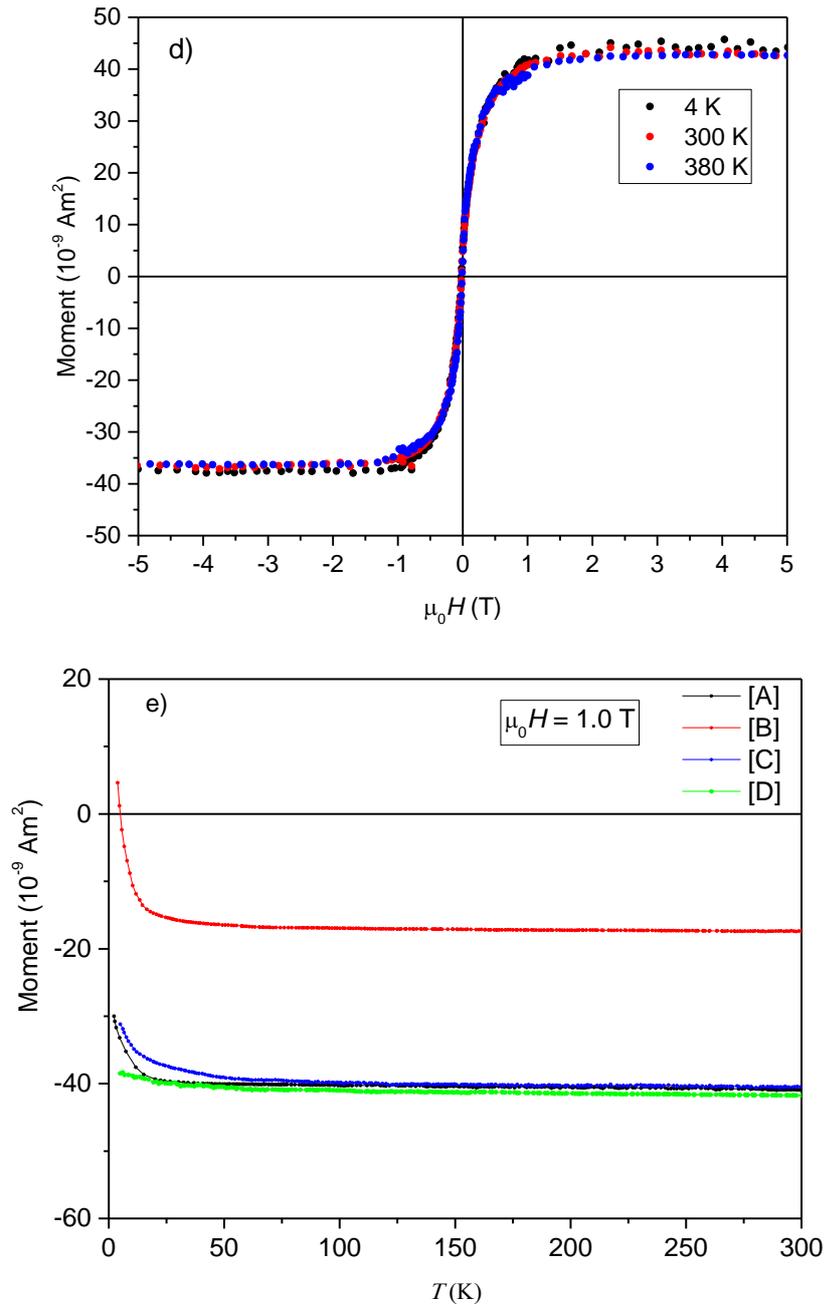

Figure 2. Some representative magnetization signals: a) As measured magnetization of sample 100 (A), as received crystal, crystal after vacuum annealing, powder and powder after vacuum annealing. b) Data corrected for the background diamagnetism of SrTiO$_3$ and the gelcap used to hold the powder. c) Temperature dependent, hysteretic magnetization curves of a 110 [A] crystal. d) Temperature dependence of magnetization of 100 [A] annealed crystal powders and e) Temperature-dependent susceptibility of 100 crystals from four different suppliers.

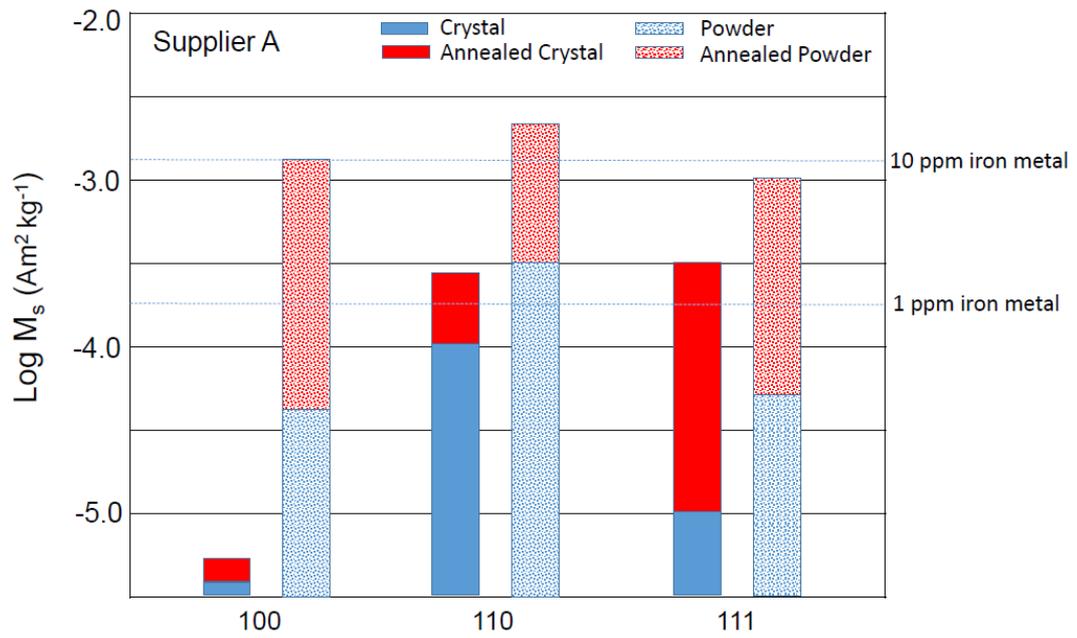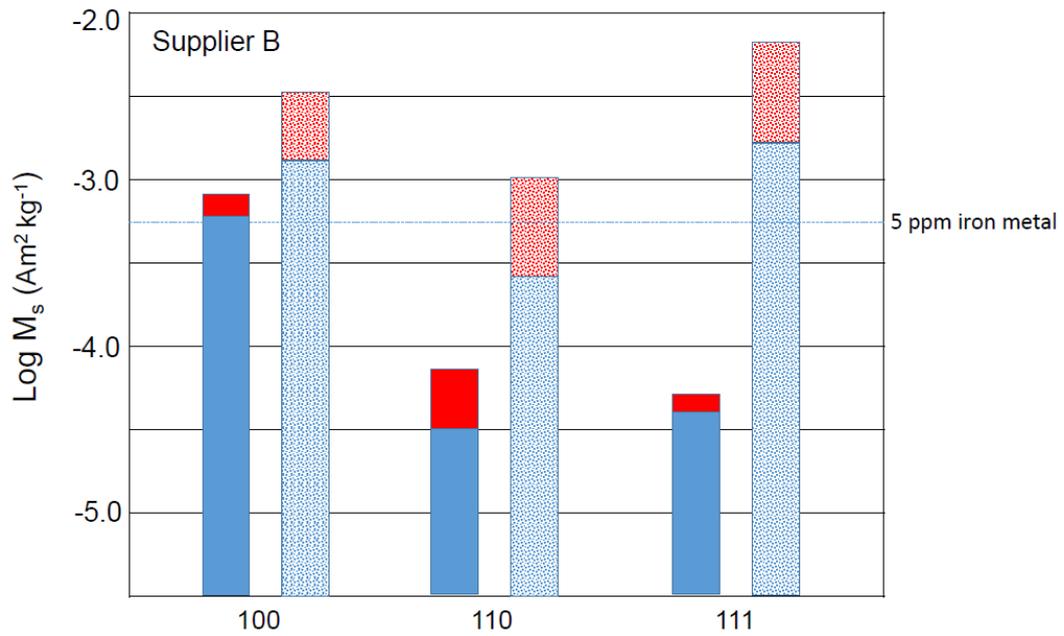

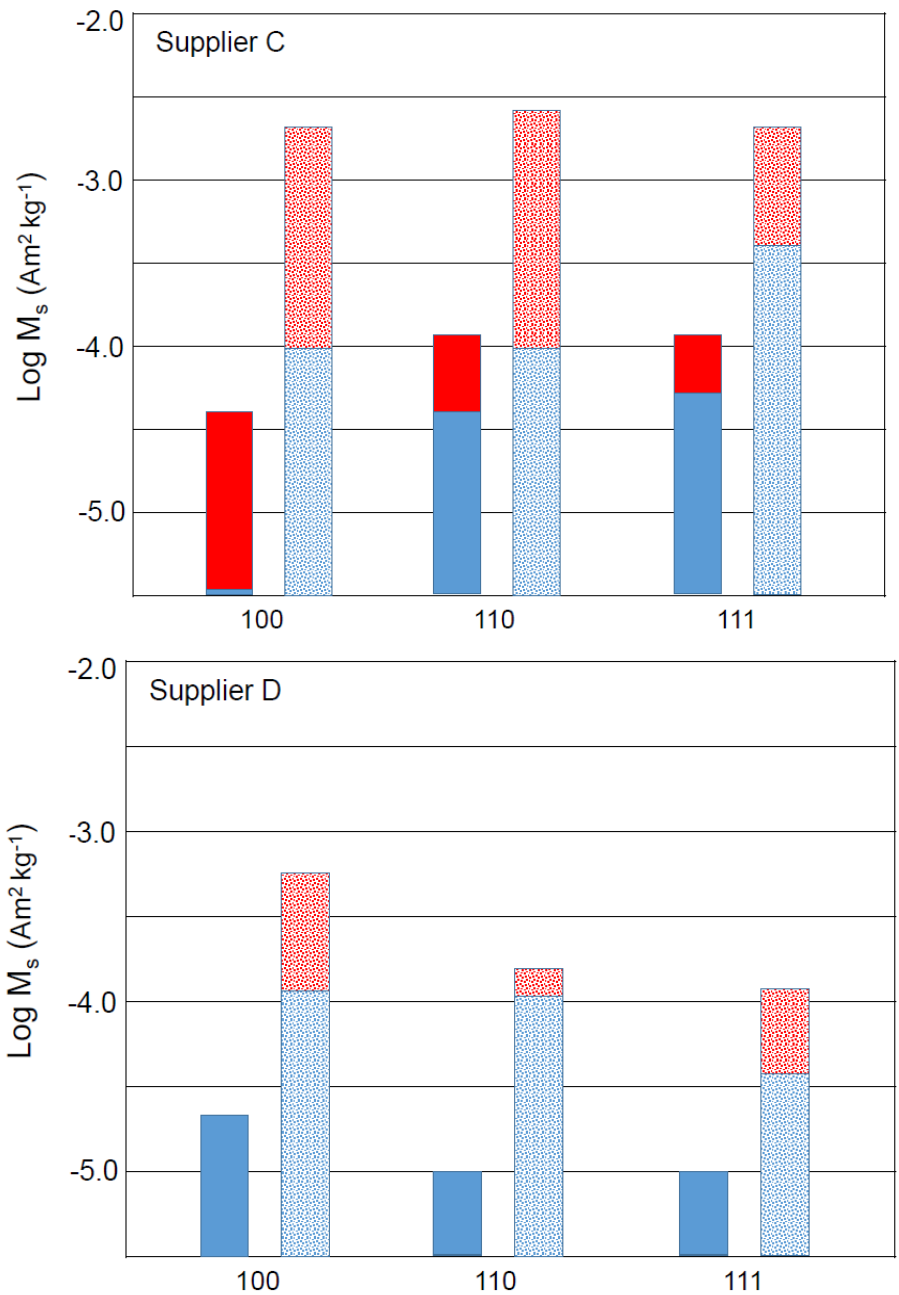

Figure 3 Summary of specific magnetization for 100, 110 and 111 crystals from four different suppliers: Blue and red bars are for samples before and after heating in vacuum at 750° C for 2 hrs. Solid bars are for single crystals, speckled bars are for corresponding powders. Note that the data are plotted on a log scale. The dashed lines correspond to the signal levels expected for 1 ppm and 10 ppm of iron metal impurity. All iron contents for 100 crystals are < 0.5 ppm, except for Supplier 'B', where it is 5 ppm.

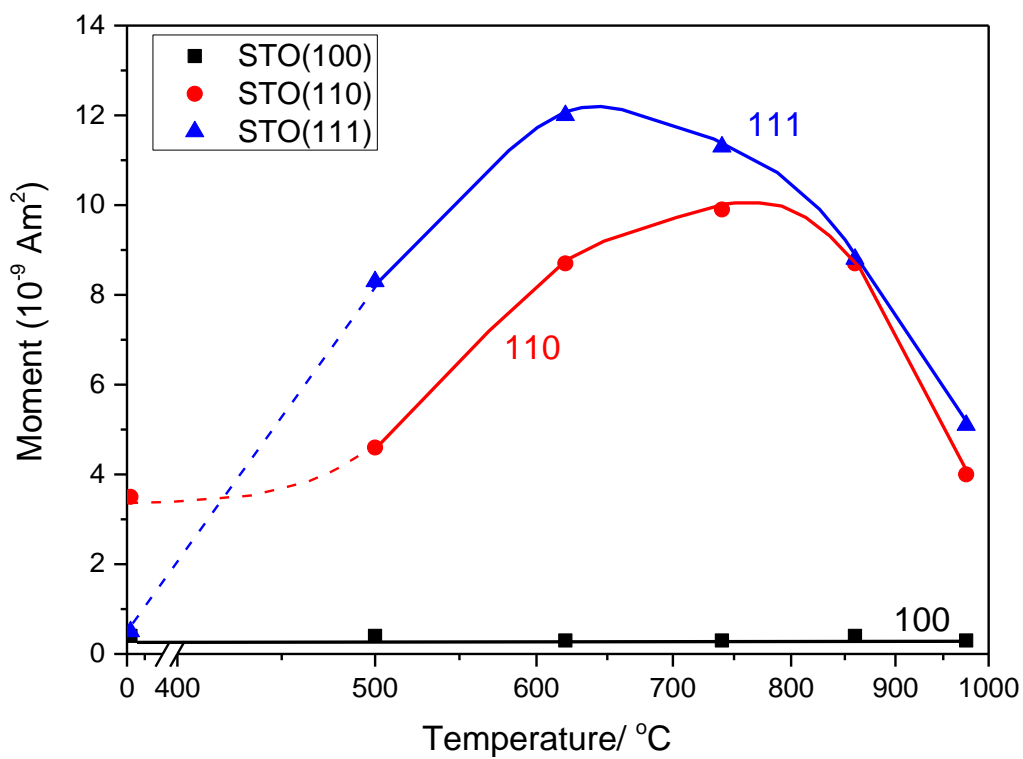

Figure 4 Dependences of crystal magnetic moments on annealing in vacuum: Data are for 100, 110 and 111 crystals from supplier A. The largest magnetic moment is found when 110 or 111 crystals are vacuum-annealed at 600 – 750 °C. No moment appears at any temperature for 100 crystals

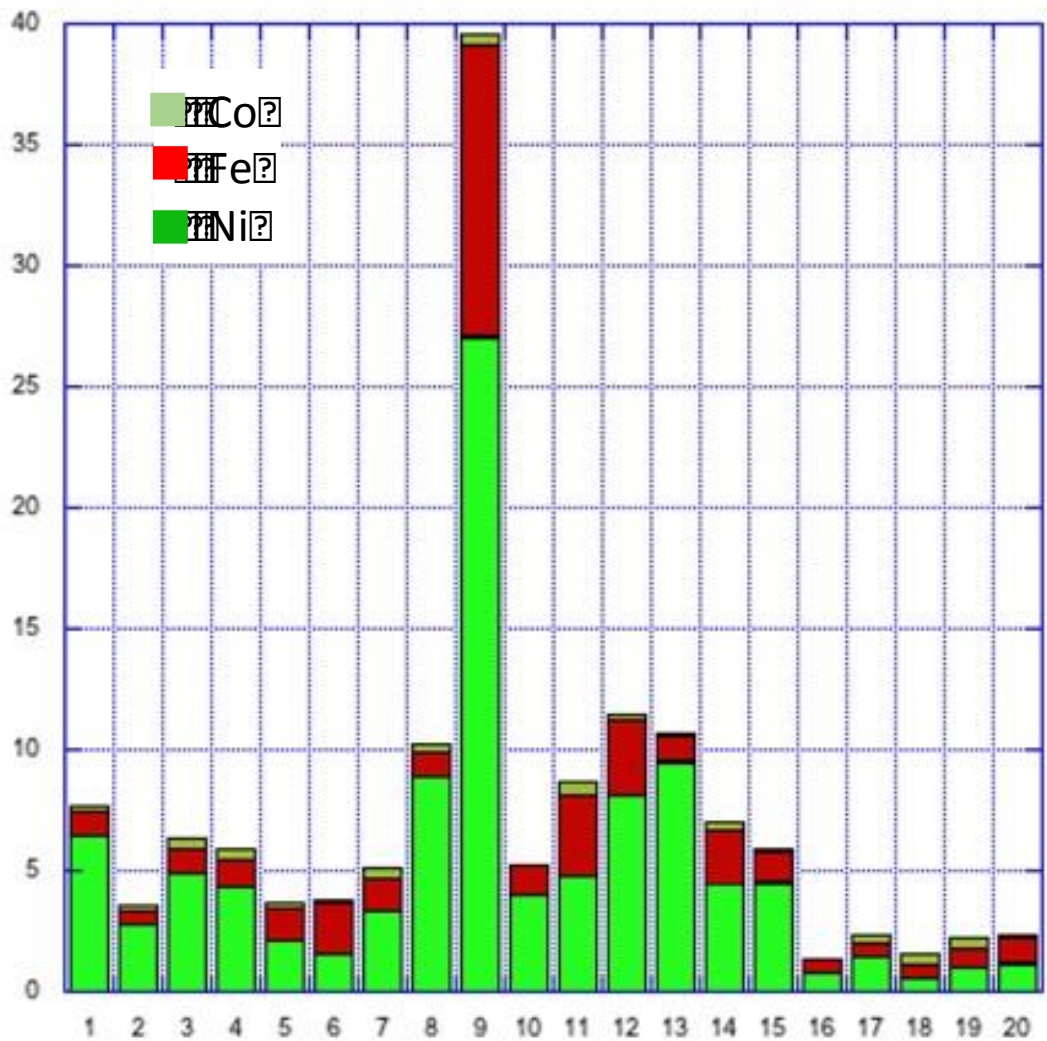

Fig 5  Surface chemical analysis scans. Concentrations of Fe, Co and Ni measured across the surfaces of 100 [A] sample by laser ablation mass spectroscopy. The distribution is very non-uniform, suggesting that particles of a transition-meta-rich impurity phase are embedded in the surface.

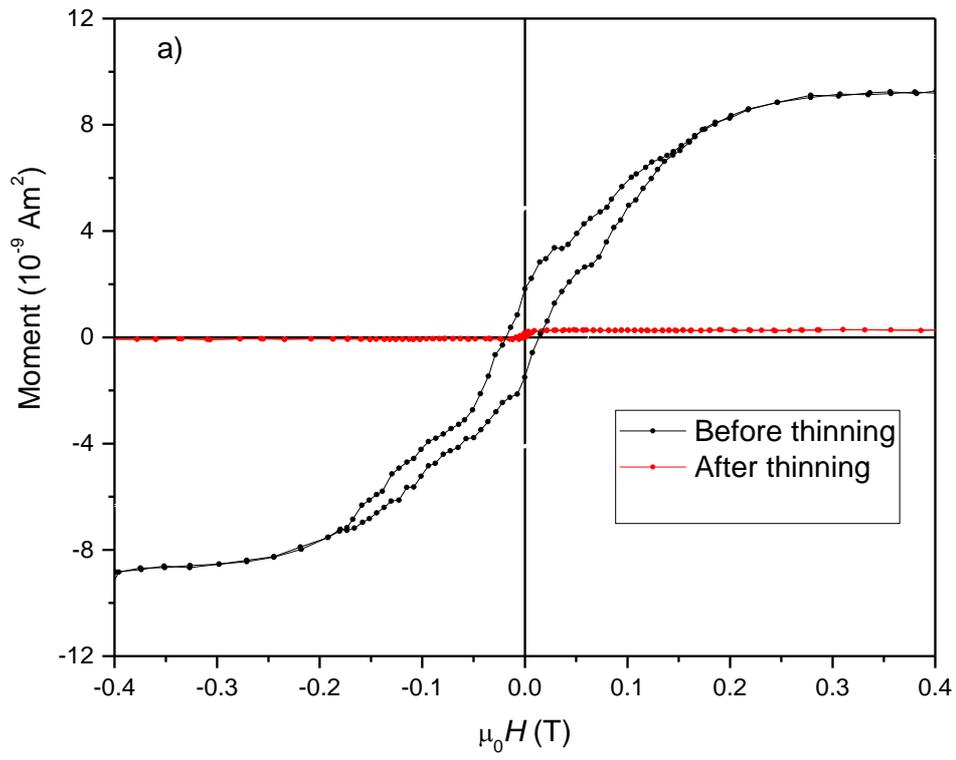

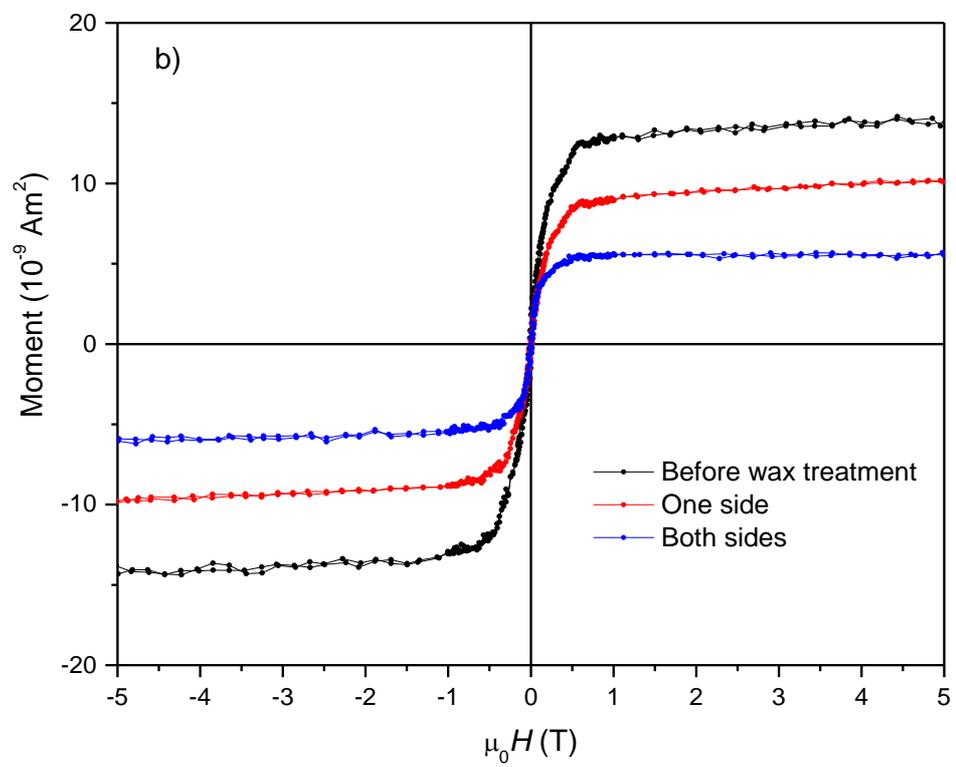

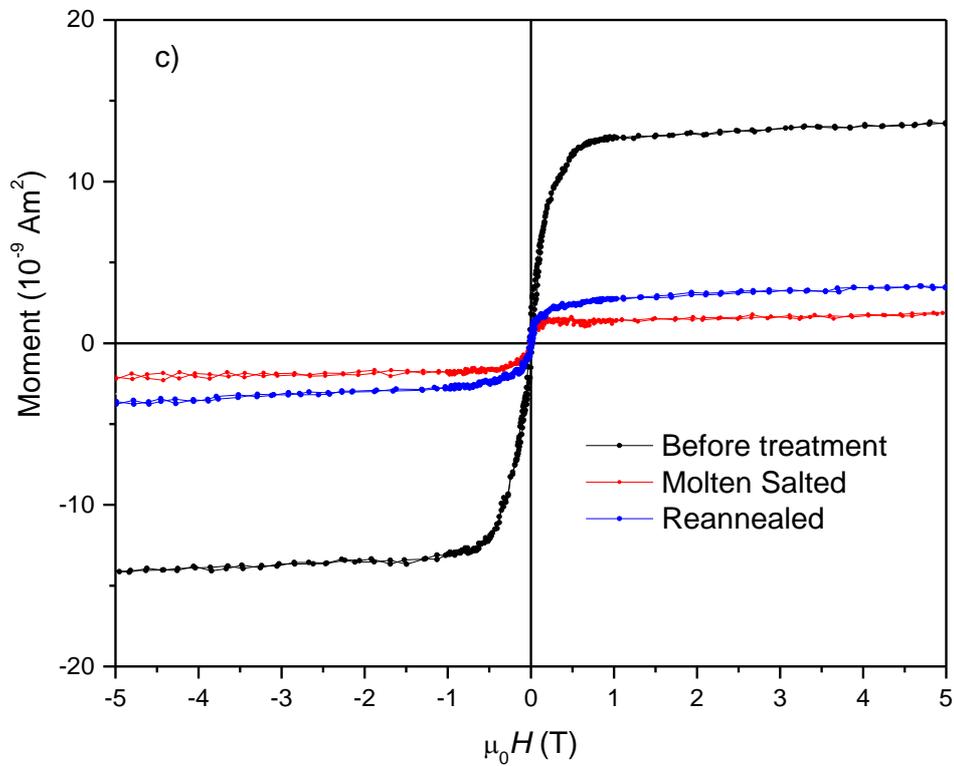

Fig 6. Variation of the magnetic moment of 110[A] crystals annealed at 750°C after surface treatment. a) mechanical thinning from 0.5 mm to 0.15 mm b) hot wax treatment at 150°C on one or both sides c) treatment in molten NaCl at 900°C with the effect of reannealing at 750°C.

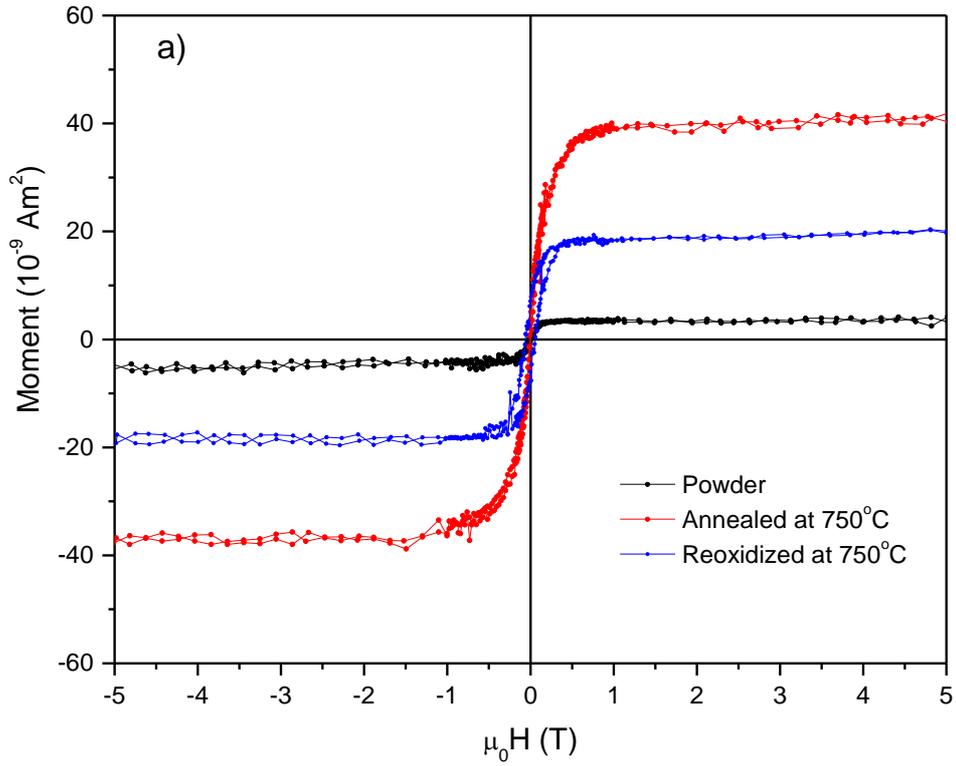

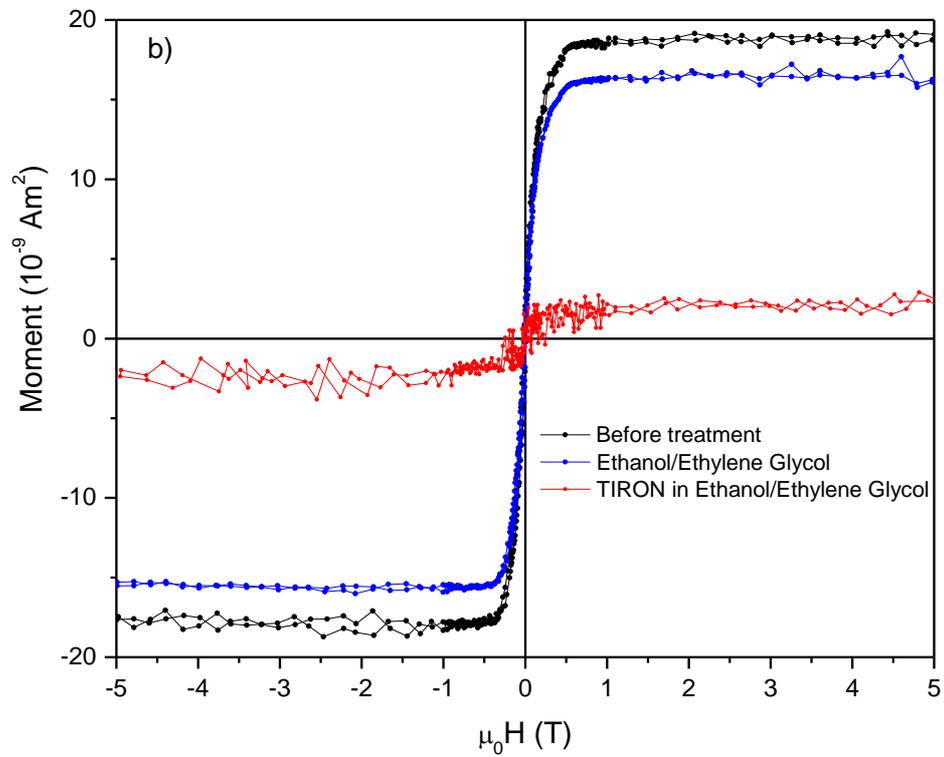

Fig 7. Treatments of annealed SrTiO₃ single-crystal powders: Effects of a) annealing 111 (C) powder in vacuum and reannealing the reduced powder in oxygen b) immersion of the reduced powder in ethylene glycol and TIRON.

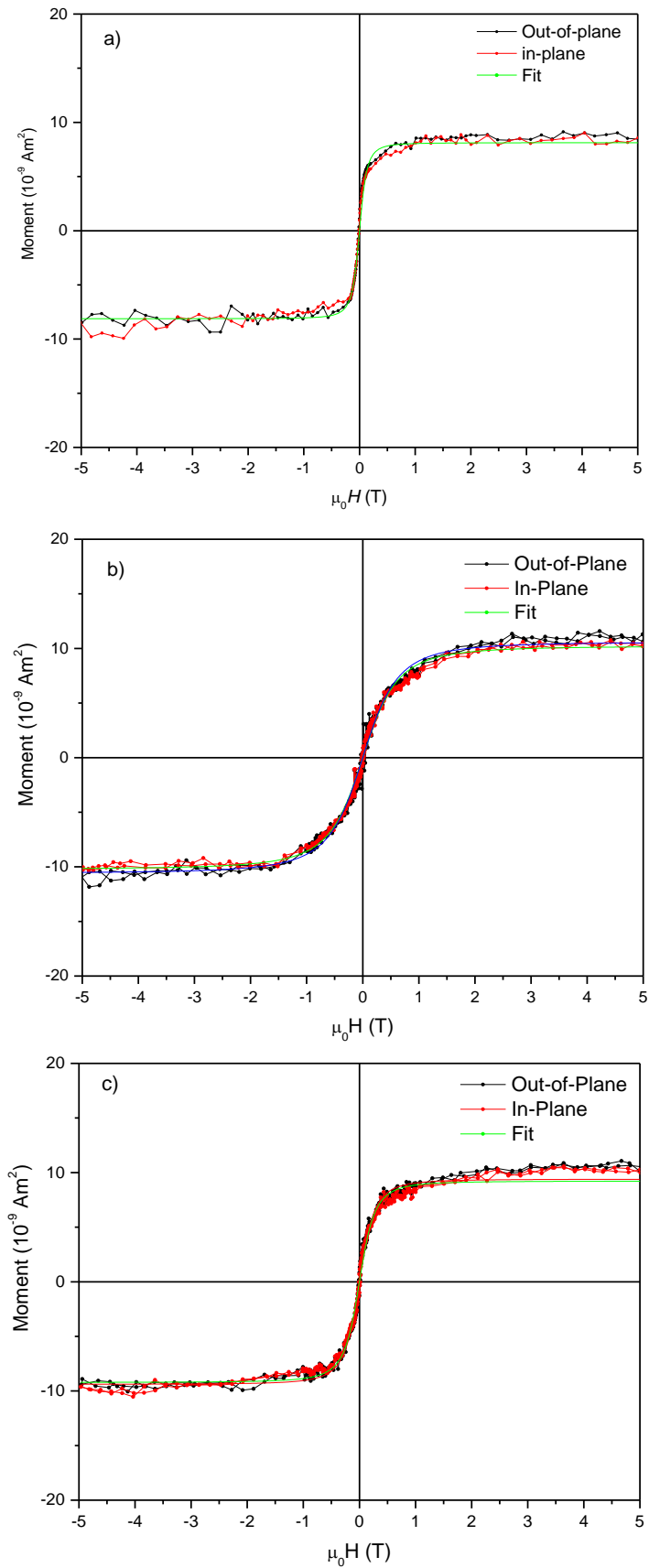

Figure 8 Magnetization curves of SrTiO$_3$ crystals as a function of applied field direction: a) 110 [A]; b) 110 [C]; c) 111 [C]

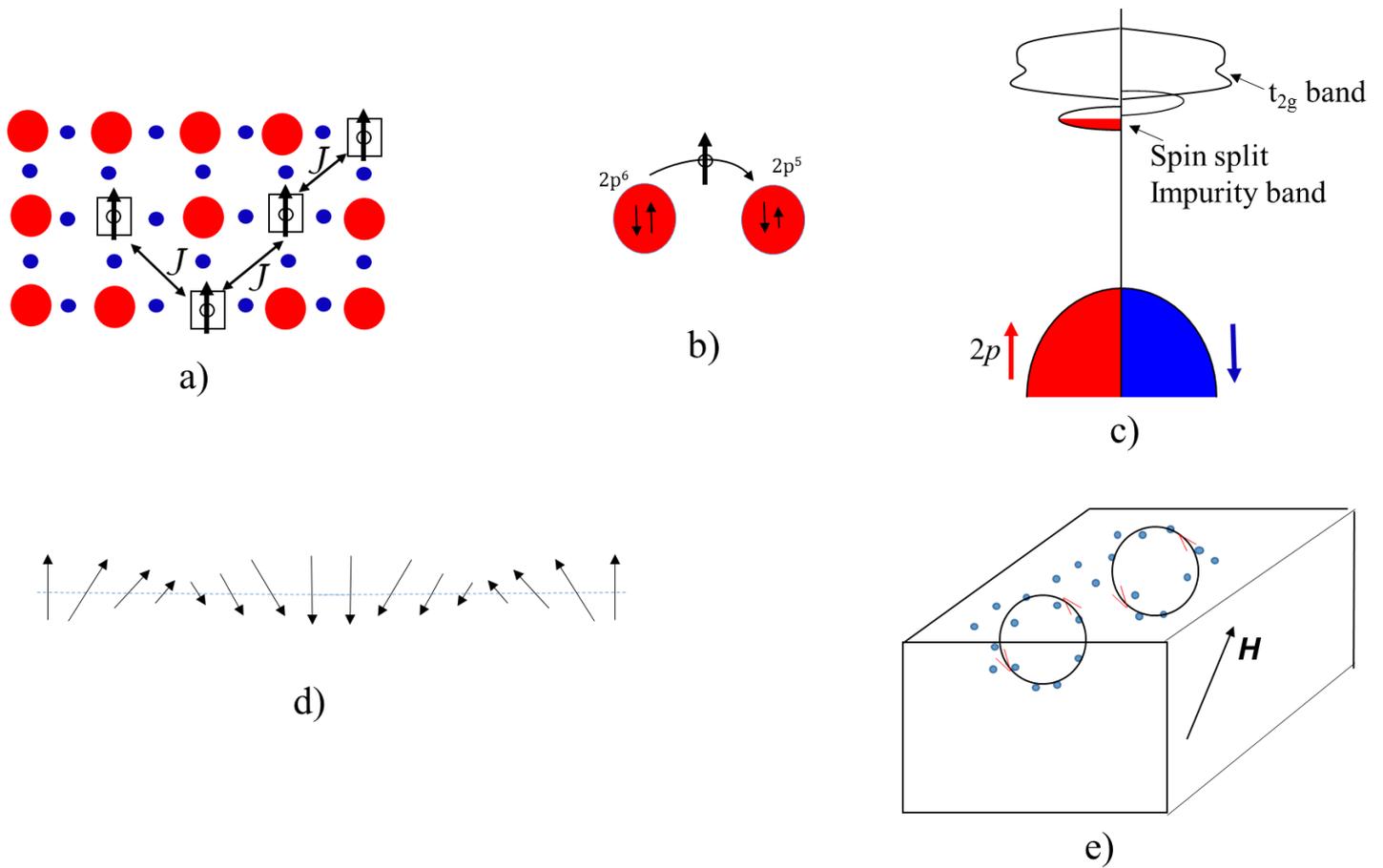

Figure 9. Schematic depiction of models considered to explain the high-temperature anhysteretic magnetic signal originating from the SrTiO$_3$ surface: a) Heisenberg ferromagnetism of electrons trapped at oxygen vacancies; b) Zener ferromagnetism of O$^-$ ions due to hole hopping in a 2p band c) Stoner ferromagnetism of a spin-split defect-related impurity band and d) giant orbital paramagnetism. The five models are discussed in the text, and summarized in Table 3.